\documentclass[10pt,letterpaper,journal, 10pt, twoside]{IEEEtran}
\usepackage{array}
\usepackage{verbatim}
\usepackage{multirow}
\usepackage{amsmath}
\usepackage{amssymb}
\usepackage{graphicx}
\usepackage[unicode=true,
 bookmarks=true,bookmarksnumbered=false,bookmarksopen=false,
 breaklinks=false,pdfborder={0 0 0},backref=false,colorlinks=false]
 {hyperref}

\makeatletter

\pdfpageheight\paperheight
\pdfpagewidth\paperwidth

\providecommand{\tabularnewline}{\\}

\ifCLASSINFOpdf
\else
\fi

\usepackage{url}
\usepackage{relsize}

\usepackage[noadjust, sort]{cite}
\usepackage{multirow}

\makeatother

\begin{document}

\title{Parallel Interleaver Design for a High Throughput HSPA+/LTE Multi-Standard
Turbo Decoder}

\author{Guohui~Wang,~\IEEEmembership{Student Member,~IEEE,} 
Hao~Shen,
Yang~Sun,~\IEEEmembership{Member,~IEEE,} 
Joseph~R.~Cavallaro,~\IEEEmembership{Senior Member,~IEEE,}
Aida~Vosoughi,~\IEEEmembership{Student Member,~IEEE,} 
and~Yuanbin~Guo,~\IEEEmembership{Senior Member,~IEEE}

\thanks{Manuscript received September 14, 2013; revised November 28, 2013, February 01, 2014; accepted February 02, 2014. 
This paper was presented in part at the IEEE International Symposium on Circuits and Systems (ISCAS), Beijing, China, May 2013~\cite{wang:iscas2013:parallel_interleaver}. This work was supported in part by Huawei, and by the US National Science Foundation under grants CNS-1265332, ECCS-1232274, and EECS-0925942.}

\thanks{G.~Wang, H.~Shen, Y.~Sun, J.~R.~Cavallaro and A.~Vosoughi are with the Department of Electrical and Computer Engineering, Rice University, Houston, TX 77005, USA (email: wgh@rice.edu). Y.~Guo is with Wireless R\&D, US Research Center, Futurewei Technologies, 5340 Legacy Dr., Plano, Texas 75024, USA.}
\thanks{Copyright (c) 2014 IEEE. Personal use of this material is permitted. However, permission to use this material for any other purposes must be obtained from the IEEE by sending an email to pubs-permissions@ieee.org.}
\thanks{Digital Object Identifier 10.1109/TCSI.2014.2309810}
}

\maketitle

%, VOL. 61, NO. 5, MAY 2014
\markboth{IEEE TRANSACTIONS ON CIRCUITS AND SYSTEMS---I: REGULAR PAPERS}{Wang \MakeLowercase{et al.}: Parallel Interleaver for High Throughput HSPA+/LTE Turbo Decoder}

\begin{abstract}
To meet the evolving data rate requirements of emerging wireless communication
technologies, many parallel architectures have been proposed to implement
high throughput turbo decoders. However, concurrent memory reading/writing
in parallel turbo decoding architectures leads to severe memory conflict
problem, which has become a major bottleneck for high throughput turbo
decoders. In this paper, we propose a flexible and efficient VLSI
architecture to solve the memory conflict problem for highly parallel
turbo decoders targeting multi-standard 3G/4G wireless communication
systems. To demonstrate the effectiveness of the proposed parallel
interleaver architecture, we implemented an HSPA+/LTE/LTE-Advanced
multi-standard turbo decoder with a 45nm CMOS technology. The implemented
turbo decoder consists of 16 Radix-4 MAP decoder cores, and the chip
core area is 2.43\,$mm^{2}$. When clocked at 600\,MHz, this turbo
decoder can achieve a maximum decoding throughput of 826\,Mbps in
the HSPA+ mode and 1.67\,Gbps in the LTE/LTE-Advanced mode, exceeding
the peak data rate requirements for both standards. 

\end{abstract}

%\IEEEpeerreviewmaketitle

\section{Introduction\label{sec:Introduction}}

\IEEEPARstart{D}{uring} the past few years, modern 3G/4G wireless
communication systems such as 3GPP (3rd Generation Partnership Project)
UMTS/HSPA+ (Universal Mobile Telecommunications System/High-Speed
Packet Access Evolution)~\cite{3GPP_HSPA+}, 3GPP LTE (Long Term
Evolution) and LTE-Advanced~\cite{3GPP_LTE} have been deployed to
meet the ever-growing demand for higher data rates and better quality
of service. High throughput is one of the most important requirements
for emerging wireless communication standards. For instance, the 3GPP
UMTS standard Release 11 extends HSPA+ with several key enhancements
including increased bandwidth and number of antennas. These enhancements
lead to 336\,Mbps peak data rate with $2\times{}2$ MIMO (multiple-input
multiple-output) and 40MHz bandwidth (or $4\times{}4$ MIMO and 20MHz
bandwidth)~\cite{3GPP_HSPA+}. Recently, up to 672\,Mbps data rate
has been proposed for the future release of 3GPP standards~\cite{Qualcomm:hspa+_advanced,maternia:ICC2012:long_term_hspa}.
As a 4G candidate, the 3GPP LTE-Advanced promises up to 1~Gbps data
rate as its long term goal.

Turbo codes are specified in many wireless communication standards
such as the HSPA+ and LTE/LTE-Advanced as forward error-correction
codes to ensure reliable communications via wireless channels, due
to their outstanding error-correcting performance~\cite{Berrou:turbo:ICC1993}.
A turbo decoder contains two key components: soft-input soft-output
(SISO) decoders and interleavers. During the decoding process, log-likelihood
ratio (LLR) soft values are exchanged between component SISO decoders
in an iterative way. The interleaver is a critical component for the
turbo decoder to achieve good error-correcting performance, by permuting
the LLRs randomly between iterations and maximizing the effective
free distance of turbo codes. Since the introduction of turbo codes,
numerous VLSI architectures and implementations have been proposed~\cite{ZFWang:TVLSI2002:area_efficient,bougard:ISSCC2003:XMAP,Benkeser:2009:JSSC,Ilnseher:2010:ICCS}.
To achieve high throughput, parallel turbo decoding architectures
are usually employed, in which several SISO decoders operate in parallel
with each working on a segment of the received codeword~\cite{Thul:2005:JVLSI,salmela:EUROPSP2007:paralle_memory_organization,martina:TCASII2008:UMTS-WiMax,Benkeser:2009:JSSC,Ilnseher:2010:ICCS,May:date2010:turbo,Wong:interleaver:jssc2010,Wang:turbo_interleaver_asap2011,Studer:2011:JSSC,lin:VLSI2011:multistandard,sun2011efficient,Asghar_interleaver_JSPS_2012,Sani:TSP2013:memory_mapping,Wang:2007:TCASII,murugappa:DATE2013:turbo,Belfanti:VLSI2013:1GTurbo}.
The parallel turbo decoders suffer from severe memory conflict problems
due to the randomness of the interleaver, which becomes a major challenge
for high throughput designs and implementations~\cite{Giulietti:EL2002:interleaver,Thul:ISCAS2002-1,Speziali:interleaver:DSD2004,Tarable:TIT2004,berrou:ICC2004:arp,Thul:2005:JVLSI,Neeb:noc:ISCAS:2005,takeshita2006:TIT:QPP,Moussa:interleaver:DATE2007,Nimbalker:interleaver:TC2008,Wong:interleaver:jssc2010,Eid:interleaver:IACAS:2011,Wang:turbo_interleaver_asap2011,Asghar_interleaver_JSPS_2012,briki:GLSVLSI2012:network_based,Sani:TSP2013:memory_mapping,Nieminen:TIT2013:turbo_butterfly}.

Although LTE and LTE-Advanced can provide higher data rates than HSPA+,
the high cost of LTE infrastructure prevents the rapid deployment
of LTE systems; in contrast, existing infrastructure can be upgraded
with low cost to support the new features defined in HSPA+. Therefore,
HSPA+ and LTE are expected to co-exist for a long term, especially
when HSPA+ still keeps evolving rapidly~\cite{Qualcomm:hspa+_advanced,maternia:ICC2012:long_term_hspa}.
It is of great interest for mobile devices to support multiple standards.
As an essential building block, multi-standard turbo decoders have
been studied in the literature~\cite{park:TVLSI2007:simd,martina:TCASII2008:UMTS-WiMax,Kim:unified:CICC2009,Ilnseher:2010:ICCS,May:date2010:turbo,Eid:interleaver:IACAS:2011,Wang:turbo_interleaver_asap2011,Asghar_interleaver_JSPS_2012,lin:VLSI2011:multistandard,Lin:JSPS2013:multi-turbo,murugappa:DATE2013:turbo}.
However, since the challenging memory conflict problem caused by the
HSPA+ interleaving algorithm limits the parallelism degree of turbo
decoders, none of them can meet the high throughput requirements of
the recent HSPA+ extensions (336\,Mbps for 3GPP Release 11 and 672\,Mbps
proposed for 3GPP Release 12 and up). 

In this paper, we propose a parallel interleaving architecture to
solve the memory conflict problem for highly parallel multi-standard
turbo decoding. The remainder of the paper is organized as follows.
Section~\ref{sec:background} introduces the memory conflict problem
in parallel turbo decoding and related work focusing on the solutions
to solve the memory conflict problem. In Section~\ref{sec:balanced-scheduling},
we describe a balanced turbo decoding scheduling scheme to eliminate
memory reading conflicts. In Section~\ref{sec:Double-buffer-Contention-free-Ar},
we present a double-buffer contention-free (DBCF) buffer architecture
to solve the memory writing conflict problem. In Section~\ref{sec:Efficient-Unified-Interleaver/De},
we adapt a unified and on-the-fly interleaver/deinterleaver address
generation architecture to the multi-standard turbo decoding architecture.
In Section~\ref{sec:VLSI-Implementation-Results} we present a high
throughput implementation of HSPA+/LTE multi-standard turbo decoder
to demonstrate the effectiveness of the proposed architecture. Finally,
we conclude this paper in Section~\ref{sec:Conclusion}.

\section{Background and Challenges\label{sec:background}}

The turbo code encoder specified in HSPA+ and LTE standards consists
of two 8-state component convolutional encoders and an interleaver.
As is shown on the left hand side of Fig.\ref{fig:turbo_enc_dec},
when a block of information bits $x_{k}\:(k=0,1,...,K-1)$ is streamed
into an encoder, a sequence of systematic bits $x_{k}^{S}$ ($x_{k}^{S}=x_{k}$),
a sequence of parity-check bits $x_{k}^{P1}$ and a second sequence
of parity-check bits $x_{k}^{P2}$ are generated and transmitted over
the wireless channel. The interleaver permutes the input bits $x_{k}$
to $x_{\pi(k)}$ based on the interleaving law $\pi$ defined in standards.

\subsection{Turbo Decoding Algorithm}

\begin{figure}[t]
\begin{centering}
\includegraphics[width=1\columnwidth]{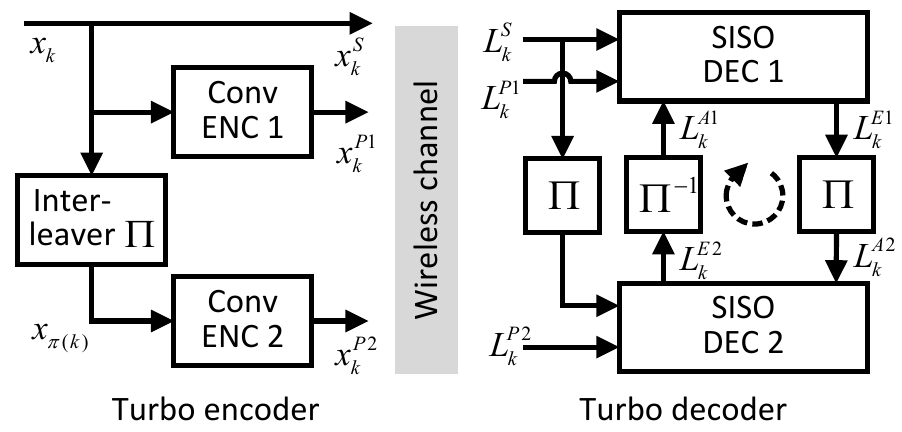}
\par\end{centering}

\vspace{-5pt}

\caption{Block diagram for a turbo encoder (left) and a turbo decoder (right). }
\label{fig:turbo_enc_dec}
\end{figure}

A turbo decoder receives soft reliability information for the transmitted
sequences $x_{k}^{S}$, $x_{k}^{P1}$ and $x_{k}^{P2}$, in the form
of log-likelihood ratios (LLRs), denoted as $L_{k}^{S}$, $L_{k}^{P1}$
and $L_{k}^{P2}$, respectively. The LLR indicates the probability
of a received bit being a binary $0$ or $1$. Shown in the right
hand side of Fig.~\ref{fig:turbo_enc_dec}, a turbo decoder consists
of two constituent soft-input soft-output (SISO) decoders and an interleaver.
The idea of the turbo decoding algorithm is to iteratively update
the probability information between two SISO decoders, separated by
an interleaver and a deinterleaver. In one iteration, two SISO decoders
process reliability information and exchange so-called extrinsic information.
The computation of each SISO decoder is called one half-iteration.
In the first half-iteration, SISO decoder 1 computes extrinsic information
$L_{k}^{E1}$ using the LLRs of the received bits $L_{k}^{S}$, parity-check
bits $L_{k}^{P1}$ and \textit{a priori} information $L_{k}^{A1}$.
$L_{k}^{A1}$ is generated by deinterleaving the extrinsic information
$L_{k}^{E2}$ from the other SISO decoder ($L_{k}^{A1}=L_{\pi^{-1}(k)}^{E2}$).
Similarly, during the second half-iteration, SISO decoder 2 generates
extrinsic information $L_{k}^{E2}$ based on $L_{\pi(k)}^{S}$, $L_{k}^{P2}$
and $L_{k}^{A2}$ ($L_{k}^{A2}=L_{\pi(k)}^{E1}$). This iterative
process continues until a preset maximum number of iterations is reached
or a stopping criterion is met. 

\begin{figure}[t]
\begin{centering}
\includegraphics[width=1\columnwidth]{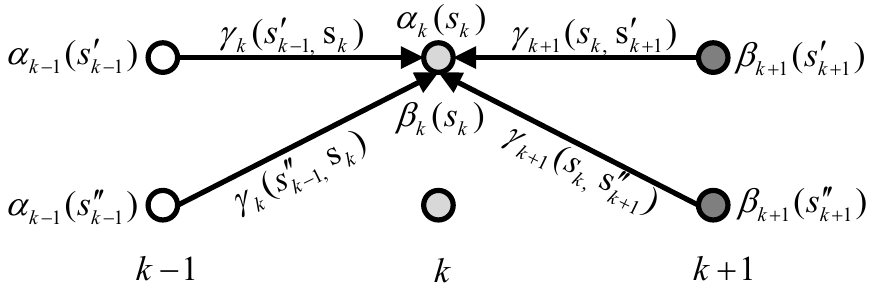}
\par\end{centering}

\vspace{-5pt}

\caption{Forward and backward recursions for trellis step $k$ in the MAP decoding
algorithm.\label{fig:Forward-and-backward}}
\end{figure}

The maximum \textit{a posteriori} (MAP) algorithm is normally used
to implement SISO decoders~\cite{Berrou:turbo:ICC1993}. The MAP
decoding process can be represented as a trellis traversal. The trellis
nodes represent the states of convolutional codes, and the branches
represent state transitions. Branch metric $\gamma_{k}(s,t)$ denotes
a transition from state $s$ to state $t$ at step $k$. As shown
in Fig.~\ref{fig:Forward-and-backward}, the trellis traversal is
performed in both forward and backward directions to compute the state
metrics $\alpha_{k}(s_{k})$ and $\beta_{k}(s_{k})$ for all eight
states: \vspace{-10pt}

\begin{align}
\alpha_{k}(s_{k})= & \overset{*}{\max}\{\alpha_{k-1}(s_{k-1}')+\gamma_{k}(s_{k-1}',s_{k}),\nonumber \\
 & \alpha_{k-1}(s_{k-1}'')+\gamma_{k}(s_{k-1}'',s_{k})\},\\
\beta_{k}(s_{k})= & \overset{*}{\max}\{\beta_{k+1}(s_{k+1}')+\gamma_{k+1}(s_{k},s_{k+1}'),\nonumber \\
 & \beta_{k+1}(s_{k+1}'')+\gamma_{k+1}(s_{k},s_{k+1}'')\},
\end{align}

\noindent{}where the $\overset{*}{\max}$ operation is typically
implemented using the max-log approximation as follows:
\begin{gather}
\overset{*}{\max}(x,y)=\max(x,y)+\log(1+e^{-|x-y|})\approx\max(x,y).
\end{gather}

Once the forward metrics and backward metrics are computed, the LLRs
of the \textit{a posteriori} probabilities (APPs) for information
bits $x_{k}$ are computed as follows:

\begin{gather}
L(\hat{x}_{k})=\overset{*}{\underset{(s_{k-1},s_{k}):x_{k}=0}{\max}}\{\alpha_{k-1}(s_{k-1})+\gamma_{k}(s_{k-1},s_{k})+\beta_{k}(s_{k})\}\nonumber \\
-\overset{*}{\underset{(s_{k-1},s_{k}):x_{k}=1}{\max}}\{\alpha_{k-1}(s_{k-1})+\gamma_{k}(s_{k-1},s_{k})+\beta_{k}(s_{k})\}.
\end{gather}

For each MAP decoder $i$ $(i\in\{1,2\})$, the extrinsic LLR information
is computed by $L_{k}^{Ei}=L(\hat{x}_{k})-L_{k}^{S}-L_{k}^{Ai}$.

\subsection{Parallel Turbo Decoding Architecture\label{sub:Parallel-Turbo-Decoding}}

To achieve high throughput, numerous parallel turbo decoding architectures
have been extensively investigated and several levels of parallelisms
have been explored~\cite{Thul:2005:JVLSI,salmela:EUROPSP2007:paralle_memory_organization,Ilnseher:2010:ICCS,May:date2010:turbo,Wong:interleaver:jssc2010,Wang:turbo_interleaver_asap2011,Studer:2011:JSSC,sun2011efficient,Asghar_interleaver_JSPS_2012,wang:iscas2013:parallel_interleaver,Sani:TSP2013:memory_mapping,murugappa:DATE2013:turbo}.

\begin{figure}[t]
\begin{centering}
\includegraphics[width=1.0\columnwidth]{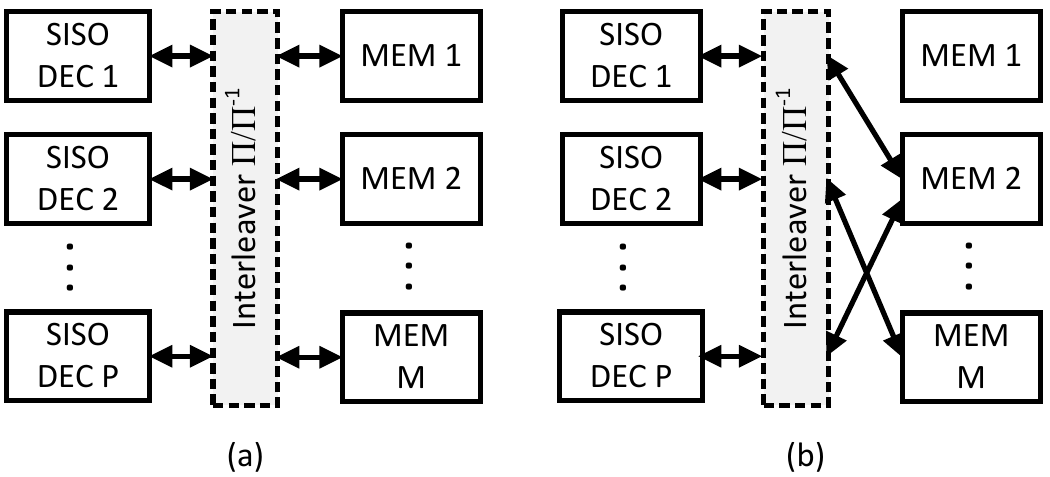}
\par\end{centering}

\vspace{-5pt}

\caption{(a) Parallel turbo decoding architecture with $P$ SISO decoders and
$M$ extrinsic memory modules; (b) A 2-way memory conflict happens
when SISO decoder 1 and SISO decoder $P$ access (read from or write
to) memory 2 at the same time.\label{fig:Parallel-turbo}}
\end{figure}

Most parallel turbo decoding algorithms exploit parallelism in SISO-decoder
level, where a codeword (with block size $K$) is partitioned into
$P$ sub-blocks with size of each being $K/P$. Multiple SISO decoders
work in parallel with each operating on one of the sub-blocks as shown
in Fig.~\ref{fig:Parallel-turbo}a. The Radix-4 SISO decoding algorithm
is another way to increase the throughput by exploring the trellis-level
parallelism. It applies a one-step look-ahead concept to the forward/backward
trellis recursions, in which two trellis steps are computed simultaneously~\cite{Kim:unified:CICC2009,sun2011efficient,Studer:2011:JSSC,Wong:interleaver:jssc2010,Asghar_interleaver_JSPS_2012}.
In addition to the above algorithms, many variants of SISO decoder
architectures explore parallelism from other aspects such as recursion-unit
level parallelism, sliding window schemes and so on~\cite{Wong:interleaver:jssc2010,Ilnseher:2010:ICCS,May:date2010:turbo,sun2011efficient,Studer:2011:JSSC,Asghar_interleaver_JSPS_2012}.
Therefore, it is quite challenging to design an interleaver architecture
which is able to handle the aforementioned different variants of parallel
turbo decoding algorithms. Our goal is to design a flexible and scalable
interleaver architecture which can fulfill this requirement.

\begin{figure}[t]
\begin{centering}
\includegraphics[width=1.0\columnwidth]{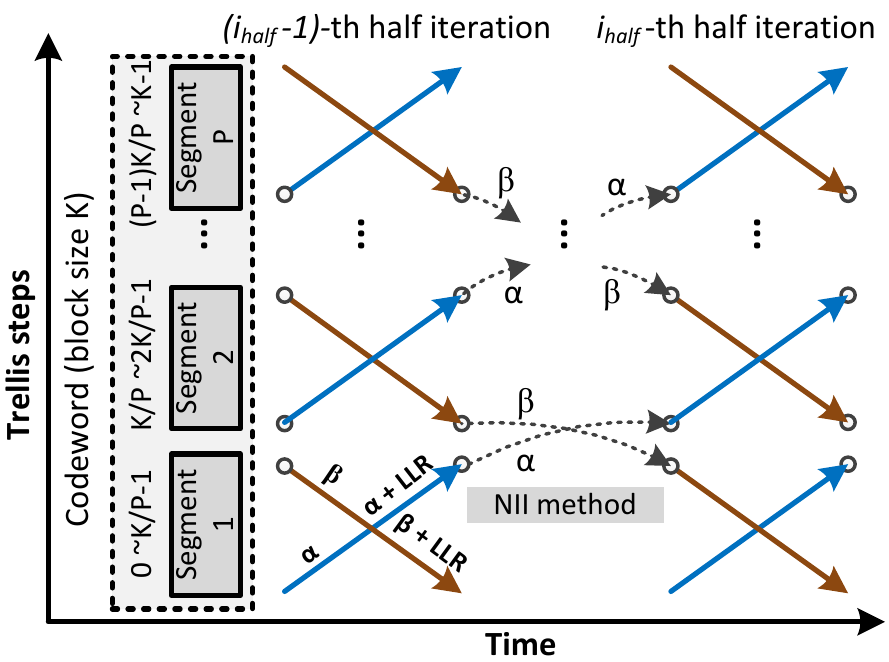}
\par\end{centering}

\vspace{-5pt}

\caption{Diagram of the cross-MAP (XMAP) decoding architecture with NII.\label{fig:xmap-dec}}
\end{figure}

In our implemented parallel turbo decoder, we choose a Radix-4 cross-MAP
(namely XMAP) decoding architecture to implement the SISO decoders
due to its high throughput performance~\cite{sun2011efficient}.
In an XMAP decoding architecture as is shown in Fig.~\ref{fig:xmap-dec},
$\alpha$ and $\beta$ recursion units work in parallel in a crossed
manner, towards the forward and backward directions, respectively.
In each half iteration, before the cross point, the $\alpha$ and
$\beta$ recursion units compute $\alpha$ and $\beta$ state metrics.
After the cross point, each Radix-4 XMAP decoder starts to produce
four extrinsic LLR values per clock cycle. When $P$ such SISO decoders
working in parallel to decode a codeword with block size $K$, each
SISO decoder performs computation on a segment (or sub-block) of the
codeword with size $K/P$. A method call Next Iteration Initialization
(NII) is employed to avoid acquisitions for the $\alpha$ and $\beta$
state metrics~\cite{bougard:ISSCC2003:XMAP,sun2011efficient}. Only
the stake information in the end of the recursion in a sub-block is
saved and propagated between SISO decoders.

\subsection{Design Challenges\label{sub:Design-Challenges}}

A parallel turbo decoder which generates $P_{LLR}$ extrinsic LLRs
in each clock cycle requires a memory bandwidth of $P_{LLR}$ memory
accesses per clock cycle. Memory used to store the extrinsic LLR values
are partitioned into $M$ ($M\geq P_{LLR})$ memory modules to provide
higher memory bandwidth. %
\begin{comment}
Most systems use $M=P$ memory modules, but some others use sub-banking
techniques to partition each memory module into $B$ sub-banks, so
that $M=B\cdot P$ memory banks are used. 
\end{comment}
However, due to the randomness of the interleaver/deinterleaver, multiple
SISO decoders may attempt to access the same memory bank concurrently,
which causes a memory conflict, as is depicted in Fig.~\ref{fig:Parallel-turbo}b.
A significant reduction in effective memory bandwidth caused by frequent
memory conflicts decreases the decoding throughput. Therefore, the
memory conflict problem has become a major bottleneck in the design
of a high throughput parallel turbo decoder~\cite{Giulietti:EL2002:interleaver,Thul:ISCAS2002-1,Speziali:interleaver:DSD2004,Tarable:TIT2004,berrou:ICC2004:arp,Thul:2005:JVLSI,Neeb:noc:ISCAS:2005,takeshita2006:TIT:QPP,Moussa:interleaver:DATE2007,Nimbalker:interleaver:TC2008,Wong:interleaver:jssc2010,Eid:interleaver:IACAS:2011,Wang:turbo_interleaver_asap2011,Asghar_interleaver_JSPS_2012,briki:GLSVLSI2012:network_based,wang:iscas2013:parallel_interleaver,Sani:TSP2013:memory_mapping}.

A traditional way of implementing the turbo code interleaver is static
memory mapping when the parallelism of turbo decoders is low. The
interleaving patterns are precomputed and stored in a look-up table
(LUT). For every decoding algorithm and parallel architecture, a unique
LUT is required since significantly different memory access patterns
are generated. If multiple standards and variable block sizes need
to be supported, memory usage for LUTs will grow drastically. This
apparently is not efficient for an embedded SoC implementation. Therefore,
on-the-fly interleaver address generation (IAG) is preferred in highly
parallel turbo decoders since it requires less memory and can be easily
parametrized to support different configurations and multiple standards.

\begin{figure*}[t]
\begin{centering}
\includegraphics[width=1\textwidth]{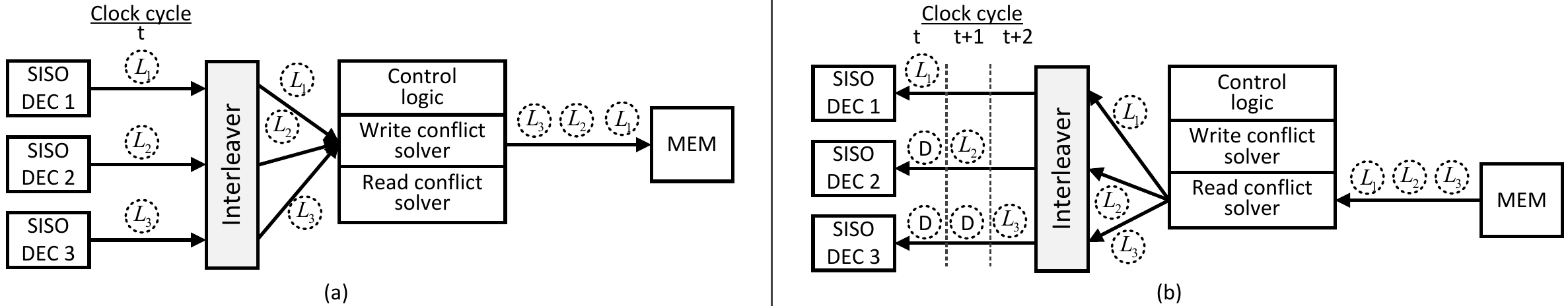}
\par\end{centering}
\begin{centering}
\vspace{-5pt}
\par\end{centering}
\caption{(a) Memory writing conflict, in which 3 SISO decoders try to write
to the same memory; (b) Memory reading conflict, in which 3 SISO decoders
try to read data from the same memory. $L_{1}$, $L_{2}$ and $L_{3}$
represent LLR data; $D$ represents a clock cycle's delay.}
\label{Fig:conflict_solver}
\end{figure*}

\subsection{Related Work}

Many interleaver architectures have been proposed to solve the memory
conflict problem in parallel turbo decoding systems, and they can
be classified into three categories: design-time solutions, compilation-stage
solutions and runtime solutions~\cite{Ilnseher:2010:ICCS,Nimbalker:interleaver:TC2008}.

(1) \emph{Design-time} \emph{solutions} usually employ algorithm-architecture
co-design methods in the early design stage~\cite{Wong:interleaver:jssc2010,Nimbalker:interleaver:TC2008}.
Contention-free algorithms are developed by applying specific algebraic
properties to the interleaver algorithms. As examples, the ARP (almost
regular permutation) interleaver adopted by the WiMax standard and
the QPP (quadratic permutation polynomial) interleaver used by LTE/LTE-Advanced
standards fall into this category~\cite{berrou:ICC2004:arp,takeshita2006:TIT:QPP}.
However, interleavers designed by these design-time methods can be
contention-free only under some constraints. For instance, the QPP
interleaver is contention-free when the decoding parallelism is a
factor of the block size $K$~\cite{takeshita2006:TIT:QPP,sun2011efficient}.
In addition, these solutions lack flexibility to support existing
interleaving algorithms, such as the 3GPP HSPA+ interleaver. Therefore,
in any of the above cases, we should seek other solutions. 

(2) \emph{Compilation-stage} \emph{solutions} employ certain memory
mapping rules to generate data access patterns to avoid memory conflicts~\cite{Giulietti:EL2002:interleaver,Tarable:TIT2004,salmela:EUROPSP2007:paralle_memory_organization,briki:GLSVLSI2012:network_based,Sani:TSP2013:memory_mapping}.
It has been shown by Tarable et al. that there is always a
memory mapping scheme which allows contention-free memory accessing
for any interleaving algorithm in a parallel turbo decoder~\cite{Tarable:TIT2004}.
Recently, Chavet et al. proposed memory mapping methods based
on graph coloring algorithms, in which a memory mapping pattern can
be calculated in polynomial time~\cite{briki:GLSVLSI2012:network_based,Sani:TSP2013:memory_mapping}.
The limitation of the compilation-stage solutions is that they require
complex off-line computations to generate memory mappings and a large
amount of memory resources to store memory mappings. To alleviate
this problem, Ilnseher et al. proposed to compress permutation
vectors to reduce the memory requirement. However, the permutation
vectors still lead to large memory area, even when a hybrid compression
method is used~\cite{Ilnseher:2010:ICCS}. 

(3) \emph{Runtime conflict} \emph{solutions} use buffers or on-chip
networks to dynamically reschedule data access orders to avoid memory
conflicts~\cite{Thul:ISCAS2002-1,Speziali:interleaver:DSD2004,Thul:2005:JVLSI,Neeb:noc:ISCAS:2005,Moussa:interleaver:DATE2007,Eid:interleaver:IACAS:2011,Wang:turbo_interleaver_asap2011,Asghar_interleaver_JSPS_2012,Nieminen:TIT2013:turbo_butterfly}.
As discussed above, both the design-time solution and compilation-stage
solution require the memory access patterns to be precomputed and
stored in memory. As a comparison, a runtime conflict solution does
not need large memories. It can provide a maximum degree of flexibility,
enabling highly configurable multi-standard turbo decoder implementation.
Therefore, we prefer runtime conflict resolution. In the rest of this
section, some existing runtime solutions are briefly described.

Thul et al. proposed a tree-based concurrent interleaving architecture
(namely TIBB) and an improved solution based on local buffer cells
interconnected via a ring network (namely RIBB), respectively~\cite{Thul:ISCAS2002-1,Thul:2005:JVLSI}.
These solutions have a high connectivity of LLR distributors, which
limit the max clock frequency of the circuits and the parallelism
of a turbo decoder. In~\cite{Speziali:interleaver:DSD2004}, Speziali
et al. extended Thul's work and proposed an improved architecture
for the interconnection network. A stalling mechanism is introduced
to the interleaver architecture. However, the stalling mechanism leads
to a high hardware complexity and results in unacceptable delay penalty
for highly parallel SISO decoder architectures such as Radix-4 SISO
decoders. Recently, Network-on-chip (NoC) approaches have been proposed
to solve the memory conflict problem. For example, in~\cite{Neeb:noc:ISCAS:2005,Moussa:interleaver:DATE2007,Nieminen:TIT2013:turbo_butterfly},
the authors proposed packet-switched NoC approaches such as Butterfly
NoC and Benes-based NoC architectures. However, these NoC methods
suffer from a large delay penalty limiting the maximum throughput.
Furthermore, to avoid network contention, complex scheduling methods
and control logic are required to schedule the network packets. In~\cite{Asghar_interleaver_JSPS_2012},
Asghar et al. presented misalignment among memory access paths
using delay line buffers. In fact, these schemes can only alleviate
the memory conflict problem when the parallelism of turbo decoder
is low. The delay line buffers also cause long delay and increase
hardware cost. 

To overcome the limitations of the aforementioned solutions, we propose
a flexible runtime interleaver architecture to solve the memory conflict
problem efficiently, so as to enable high throughput multi-standard
turbo decoding.

\section{Eliminating Memory Reading Conflicts via Balanced Turbo Decoding
Scheduling\label{sec:balanced-scheduling}}

In a traditional turbo decoding architecture, due to the existence
of data dependencies during trellis traversal in a turbo decoder,
SISO decoders should stall until the currently requested datum is
ready. Frequent stalling of SISO decoders will significantly decrease
the decoding throughput. In order to achieve high throughput, we should
keep all the SISO decoders running at full speed without stalling.
Fig.~\ref{Fig:conflict_solver} demonstrates two cases showing traditionally
how the memory conflict problem was solved using memory conflict solvers,
when three SISO decoders are trying to access (write or read) the
same memory module. We will analyze the disadvantage of these solutions,
and then discuss our solution in Section~\ref{sub:Unbalanced-and-Balanced}.

\subsection{Properties of Memory Reading and Writing Conflicts}

For a memory writing conflict (Fig.~\ref{Fig:conflict_solver}a),
a buffer structure can be used as a ``write conflict solver'' to
cache the data accesses and resolve the memory conflicts. All the
SISO decoders can execute at full speed so that the overall high throughput
can be guaranteed. In contrast, if three SISO decoders request data
from memory in the same clock cycle (Fig.~\ref{Fig:conflict_solver}b),
only SISO decoder-1 can get the requested LLR value $L_{1}$ immediately.
The SISO decoder-2 must wait for a clock cycle to get $L_{2}$ and
SISO decoder-3 must wait for two clock cycles to get $L_{3}$. Although
buffering and data prefetching may help reduce the number of memory
reading conflicts, it is difficult to completely eliminate memory
reading conflicts in a practical hardware implementation. In addition,
the reading conflict solver and the writing conflict solver are so
different that implementing both of them utilizes more hardware resources
and requires extra multiplexing logic.

\subsection{Unbalanced and Balanced Scheduling Schemes\label{sub:Unbalanced-and-Balanced}}

\begin{figure}[t]
\begin{centering}
\includegraphics[width=1\columnwidth]{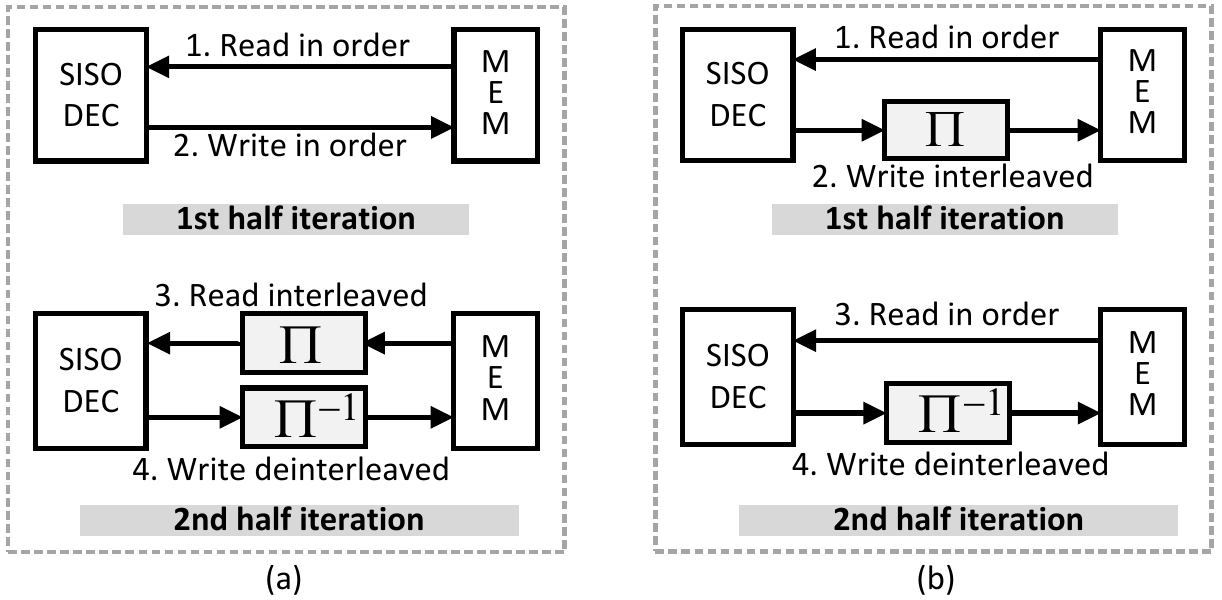}
\par\end{centering}

\begin{centering}
\vspace{-5pt}
\par\end{centering}

\caption{(a) The unbalanced turbo decoder scheduling; (b) The balanced scheduling
scheme. (The numbers indicate the order of scheduling steps.)}

\label{Fig:scheduling}
\end{figure}

In this Section, we describe the advantages of a balanced scheduling
scheme to deal with the memory conflict problem. There are two scheduling
schemes when implementing a turbo decoder, namely an unbalanced scheme
and a balanced scheme. As shown in Fig.~\ref{Fig:scheduling}a, in
an unbalanced scheduling scheme, the SISO decoder reads (writes) contiguous
data from (to) the memory in a natural order in the 1st half iteration.
During the 2nd half iteration, the SISO decoder first reads data from
memory in an interleaved way; once computation is done, the SISO decoder
writes the output data into memory in a deinterleaved way. Due to
the randomness of the interleaving/deinterleaving algorithms, both
memory reading and writing in the 2nd half iteration suffer from memory
contentions in a parallel turbo decoder. 

Fig.~\ref{Fig:scheduling}b shows the balanced scheduling scheme
in which SISO decoders write output data in the interleaved/deinterleaved
order in both half iterations, so that in the next half iteration
they can read the data in a naturally continuous order. Since all
the memory reading operations are in-order, there is no memory conflict
at all~\cite{wang:iscas2013:parallel_interleaver}. The balanced
scheduling scheme can lead to more efficient designs than the unbalanced
scheduling scheme because of the following reasons. First of all,
the memory writing conflict is easier to solve than the memory reading
conflict as is discussed in Fig.~\ref{Fig:conflict_solver}. Secondly,
without memory reading conflicts, we can remove the reading conflict
solver hardware from the interleaver and reduce the hardware cost.
Thirdly, the balanced scheduling method generates a more balanced
workload between two half iterations. Since both the read and write
conflict solvers introduce extra penalty clock cycles, the memory
conflicts occurring in both memory reading and writing in a unbalanced
scheduling scheme lead to more penalty clock cycles in the 2nd half
iteration. Furthermore, during the 2nd half iteration in unbalanced
scheme, the control logic need to keep track of the execution status
of SISO decoders, since the SISO decoders may run/stall independently.
The bookkeeping and frequent stalling/restarting operations on the
SISO decoders require more complex control logic in the unbalanced
scheduling scheme. In the balanced scheduling scheme, the symmetric
structure between two half iterations results in a more balanced workload,
which simplifies the control logic. 

Due to the above-mentioned benefits, we choose the balanced scheduling
scheme in our multi-standard turbo decoder implementation.

\subsection{Further Discussions }

Although the balanced scheduling scheme requires both interleaver
and deinterleaver hardware modules, the fact that the interleaver
and deinterleaver algorithms typically have symmetric structures implies
that hardware sharing is possible to save computational resources
and storage resources. Therefore, the overhead of a two-mode interleaver/deinterleaver
would be small compared to a single-mode interleaver, if hardware
sharing is explored. In addition, as an enabler for highly parallel
turbo decoders supporting standards that do not employ contention-free
interleaver algorithms (such as the HSPA+ standard), a small hardware
overhead can be acceptable. This hardware sharing method will be further
discussed in Section~\ref{sec:Efficient-Unified-Interleaver/De}
and Section~\ref{sub:hspa_intl_implementation_results}.

\section{Double-buffer Contention-free Architecture for Interconnection Network\label{sec:Double-buffer-Contention-free-Ar}}

By using the balanced turbo decoding scheduling scheme, we eliminate
the memory reading conflicts, but the memory writing conflicts still
exist. In this section, we describe a double-buffer contention-free
(DBCF) architecture to solve the memory writing conflict problem with
low hardware complexity~\cite{Wang:turbo_interleaver_asap2011}.
We first introduce the methodology and system model to analyze properties
of memory conflicts and then describe the DBCF architecture in detail.

\subsection{Methodology and System Model}

\begin{figure}[t]
\begin{centering}
\includegraphics[width=1\columnwidth]{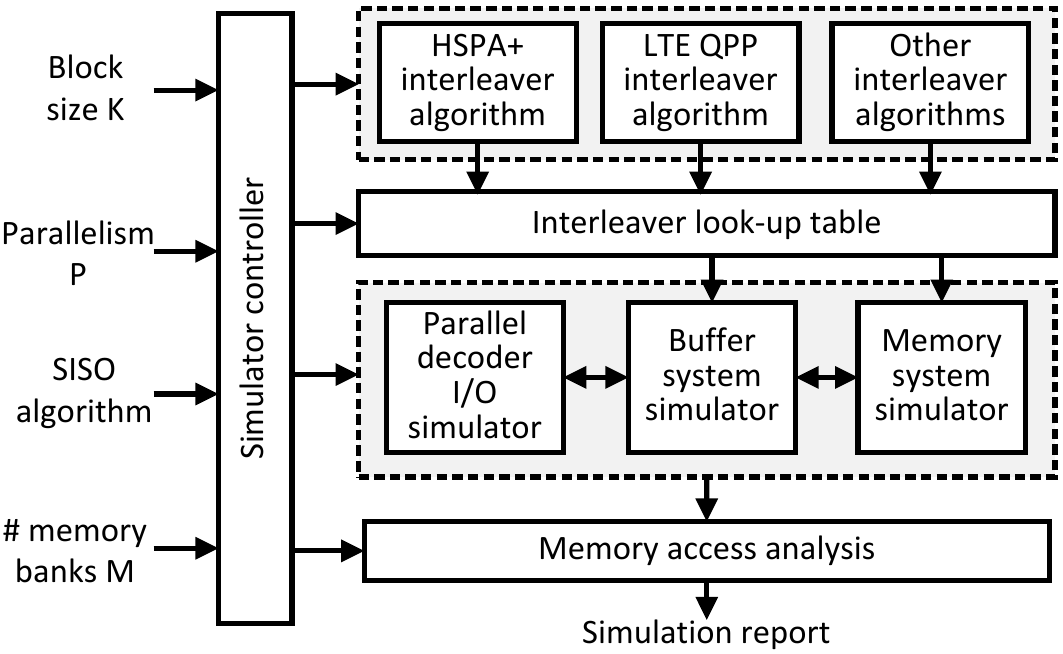}
\par\end{centering}

\vspace{-5pt}

\caption{Cycle-accurate simulation used to study the memory conflict problem.\label{fig:methodology}}
\end{figure}

To study the statistical property of memory conflicts under an interleaving
law with a special parallel architecture, a cycle-accurate simulator
is designed using a methodology shown in Fig.~\ref{fig:methodology}.
The simulator is written in the C language, and can simulate the behaviors
of SISO decoders' I/O ports, the buffer system and the memory system
in each clock cycle. Interleaving address patterns are calculated
and stored in a look-up table in the beginning of a simulation. The
simulator can be easily extended to support new interleaving laws.
This cycle-accurate simulator can also be used to verify or simulate
a memory conflict solver by modifying the buffer system simulator.
This function is used to determine some architectural parameters of
the DBCF architecture in Section~\ref{sub:DBCF_arch}.

Depending on the parallel decoding strategies and memory organizations,
the memory access patterns can vary from case to case. Our implemented
parallel turbo decoder consists of $P$ parallel Radix-4 XMAP decoders
as described in Section~\ref{sub:Parallel-Turbo-Decoding}, and $M$
memory modules storing the extrinsic LLR values. In such a system,
a Radix-4 XMAP decoder produces 4 new LLR data in each clock cycle.
The total number of LLR data to be written in the extrinsic memory
per clock cycle is defined as $P_{LLR}=4P$, which can be considered
as the effective parallelism of a parallel turbo decoder.

\subsection{Statistical Analysis of Memory Conflicts}

\begin{figure}[t]
\begin{centering}
\includegraphics[width=1\columnwidth]{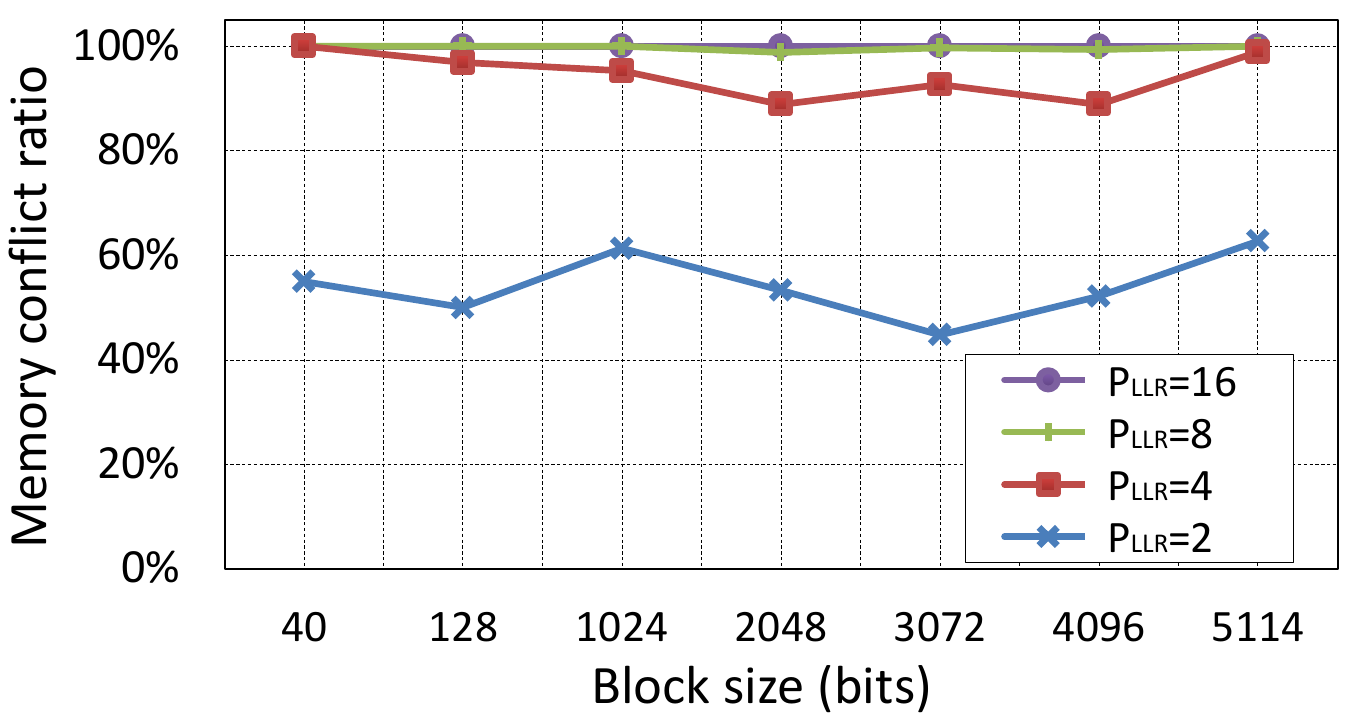}
\par\end{centering}

\vspace{-5pt}

\caption{Memory conflict ratio for different parallelism degrees and block
sizes for the HSPA+ turbo code interleaver.\label{fig:Memory-conflict-ratio}}
\end{figure}

The HSPA+ mode is the most challenging part in a high throughput multi-standard
turbo decoder due to the frequent and severe memory conflicts~\cite{Eid:interleaver:IACAS:2011,Asghar_interleaver_JSPS_2012}.
Therefore, in order to design an efficient VLSI architecture, it is
important to study the properties of memory conflicts in HSPA+ mode.
We first analyze how the memory conflict ratio changes as the block
sizes and the effective parallelism $P_{LLR}$ change by performing
cycle-accurate simulations. Memory conflict ratio (MCR) is defined
as a ratio of the number of clock cycles with memory conflicts to
the total number of clock cycles for a half iteration: $MCR=N_{conflict\_cycle}/N_{total\_cycle}$.
The MCR can indicate how severe the memory conflict problem is for
certain decoding configurations. Fig.~\ref{fig:Memory-conflict-ratio}
shows that for a fixed parallelism the MCR does not change significantly
across different block size configurations. However, when parallelism
$P_{LLR}$ goes higher, the MCR increases dramatically. For parallelism
$P_{LLR}$ higher than 8, memory conflicts occur in almost every clock
cycle, which indicates very severe memory conflict problems. Therefore,
as the parallelism goes higher, the difficulty to resolve the memory
conflict problem increases drastically.

\begin{figure}[t]
\begin{centering}
\includegraphics[width=1\columnwidth]{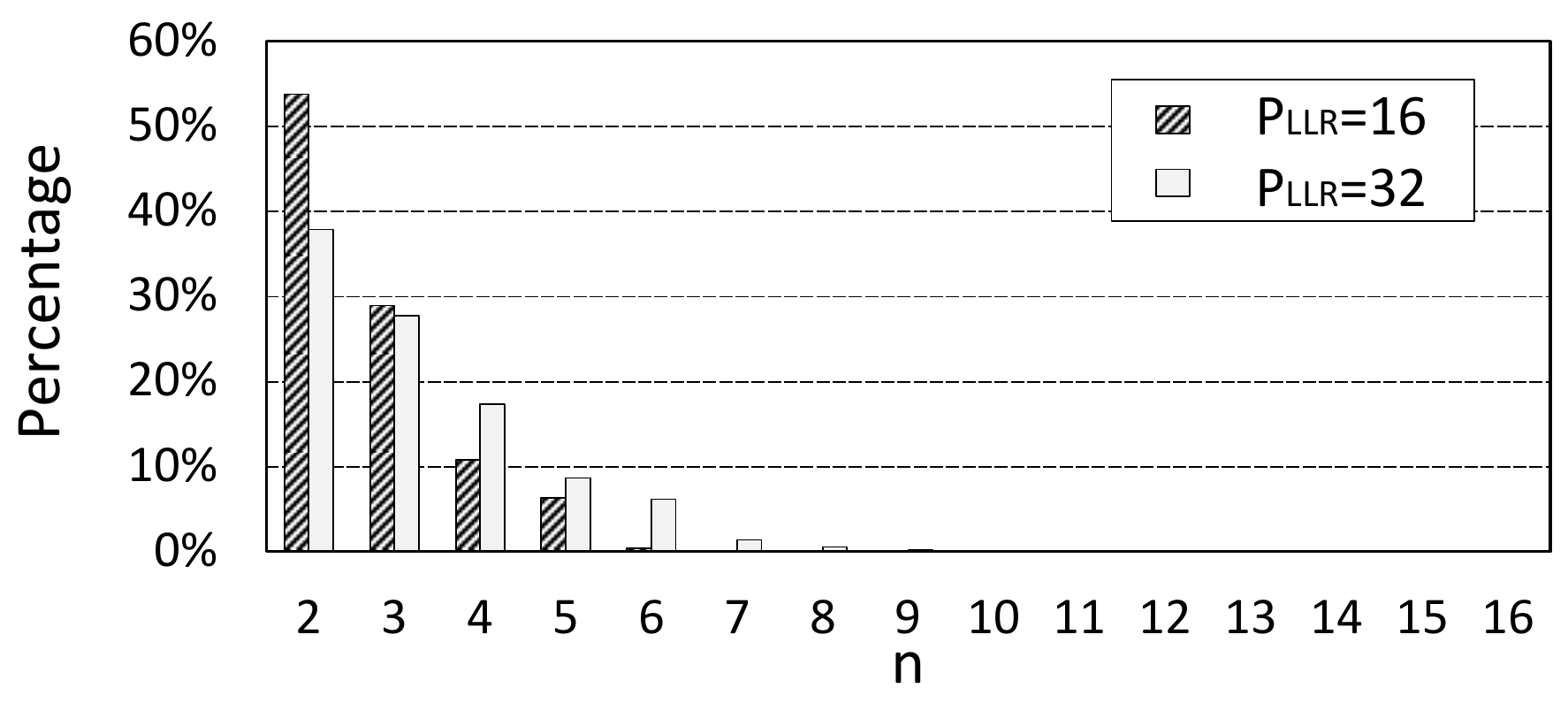}
\par\end{centering}

\vspace{-5pt}

\caption{The percentage of $n$-way memory conflicts (for $P_{LLR}=16$ and
$P_{LLR}=32$). Interleaving algorithm from 3GPP HSPA+ standard is
employed. Turbo code block size is $K=5114$.\label{fig:percentage_n_way_conflict}}
\end{figure}

Simulation results show that during the whole decoding process, the
average number of memory accesses for each memory module per clock
cycle is very close to 1. This implies the possibility to use buffers
to smooth the bursty data requests to a memory module, and finally
to achieve near 1 data/cycle/memory throughput. We further analyze
the statistical distribution of memory conflicts. If $n$ LLR data
in the same memory module are accessed by multiple SISO decoders in
one clock cycle, we call it an $n$-way memory conflict ($n\in[2,P_{LLR}])$.
The percentage of $n$-way memory conflicts ($Percent{}_{n\_way}$)
is calculated as the ratio of the number of $n$-way conflicts ($N_{n\_way}$)
to the total number of memory conflicts ($N_{Conflict}$): $Percent_{n\_way}=N_{n\_way}/N_{Conflict}$.
The percentage of $n$-way memory conflicts for parallelism of 16
and 32 in 3GPP HSPA+ interleaver are shown in Fig.~\ref{fig:percentage_n_way_conflict}.
We can notice that most of the memory conflicts are 2-, 3- and 4-way
conflicts. These three types of conflicts cover 93.35\% and 82.97\%
of the memory conflicts for $P_{LLR}=16$ and $P_{LLR}=32$, respectively.
Based on the above observations, we propose a double-buffer contention-free
architecture (DBCF) to efficiently solve the memory conflict problem.

\subsection{Double-buffer Contention-free (DBCF) Architecture\label{sub:DBCF_arch}}

\begin{figure*}[t]
\begin{centering}
\includegraphics[width=0.7\textwidth]{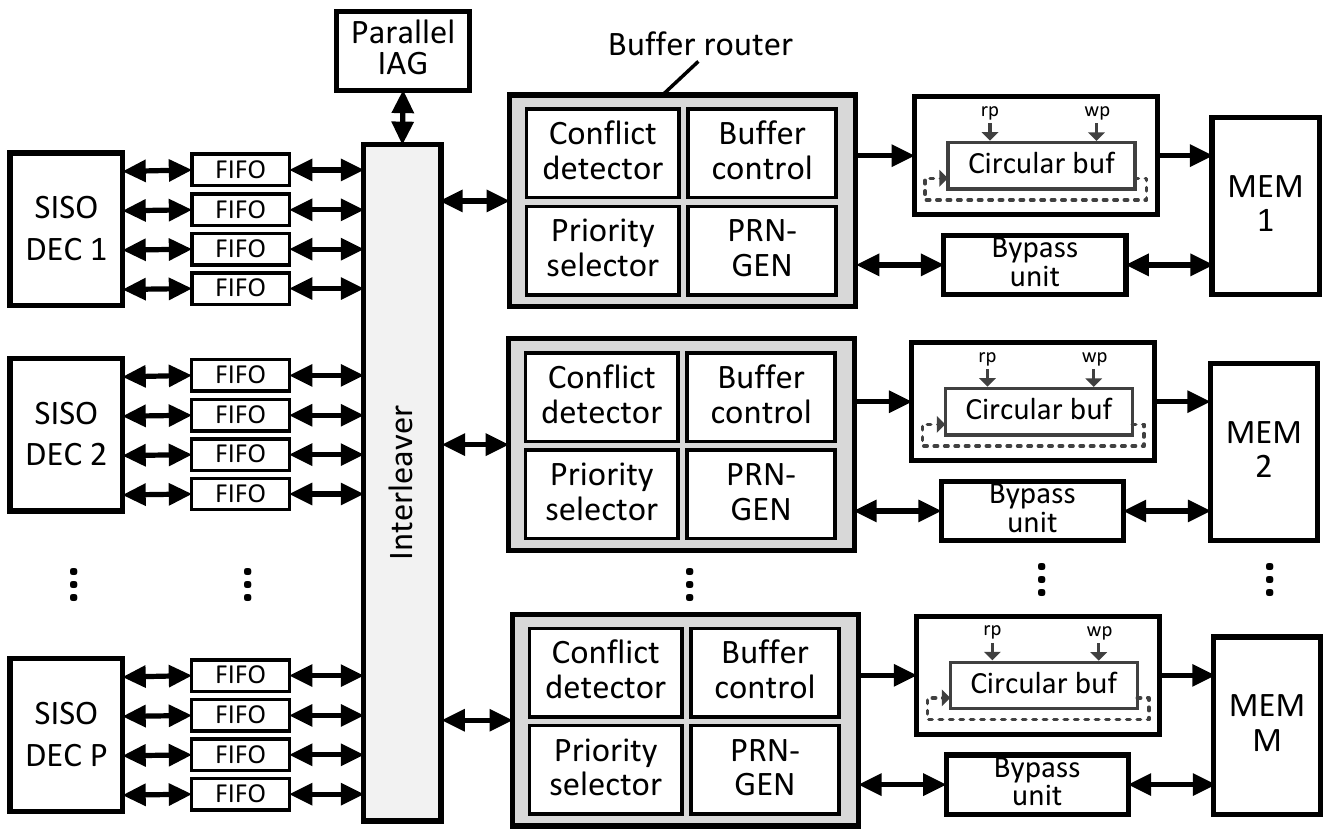}
\par\end{centering}

\caption{Block diagram of the proposed DBCF buffer architecture to the solve
memory writing conflict. (Note: wp and rp in the circular buffer represent
write pointer and read pointer, respectively.)\label{fig:DBCF} }
\end{figure*}

The DBCF buffer architecture is built around the interleaver (with
interleaving address generation units) between the SISO decoders and
memory modules. Fig.~\ref{fig:DBCF} shows the diagram of the DBCF
architecture. Major components of the DBCF architecture include FIFOs
associated with SISO decoders, circular buffer associated with memory
modules, buffer routers and bypass units. The name of DBCF architecture
comes from the fact that two sets of buffers (FIFOs and circular buffers)
are employed, which gives the proposed architecture a large advantage
over the traditional single-buffer-based architectures~\cite{Thul:2005:JVLSI,Speziali:interleaver:DSD2004}. 

\subsubsection{Micro-architecture of DBCF}

The main idea of the DBCF architecture is to use circular buffers
to cache the concurrent data writing operations. The circular buffers
should be implemented using registers in order to support concurrent
data writing. A write pointer and a read pointer are provided to control
the buffer access. If full inter-connection is fulfilled between the
interleaver and the circular buffers, the exponentially growing hardware
cost as the parallelism goes higher makes the implementation very
inefficient~\cite{Speziali:interleaver:DSD2004}. According to the
statistical analysis of memory conflicts, it is rare to have an $n$-way
memory conflict for $n>4$, therefore, a selection-based data routing
scheme is implemented by the buffer router. We define a selection
parameter $S$ ($S\in[1,N_{LLR}]$), which controls the number of
data that are allowed to be routed to the circular buffers in a clock
cycle. At any time, up to $S$ LLR data are allowed to access the
circular buffer, the other $\max(0,n-S)$ LLR data are rejected for
a $n$-way memory conflict case. For example, if there are 5 LLR data
trying to access a certain memory module ($n=5$). The conflict detector
detects a memory conflict. Assume we have set $S=3$, 3 out of 5 LLR
data will be stored into the circular buffer associated with the memory
module. At the same time, the remaining 2 LLR data ($\max(0,n-S)=\max(0,5-3)=2)$
are pushed into the FIFOs connected to the producer SISO decoders.
This selection-based data routing method reduces complexity of the
inter-connection network between the interleaver and the circular
buffers from $P_{LLR}\times P_{LLR}$ to $P_{LLR}\times S$.

The conflict detector can detect the memory conflicts at runtime using
a compare-select cluster. It is worth mentioning that in the proposed
architecture memory conflicts are detected and solved at runtime,
without any pre-stored data sets from the initial design phase. Then,
with the help of pseudo-random number generator (PRN-GEN) circuits,
priority selector circuits determine which $S$ data are selected
with equal fairness. The buffer control units maintain the writing
and reading operations to the circular buffers. The bypass units directly
route the data into the memory module when the circular buffer is
empty to avoid unnecessary latency caused by circular buffers.

The FIFOs associated with SISO decoders is one of the important reasons
why the DBCF architecture can efficiently solve the memory writing
conflict problem. If a traditional single-buffer architecture is used,
we have to stall the corresponding SISO decoders if their LLR outputs
are rejected by the buffer router circuits; otherwise, the rejected
data will be lost. Thanks to the FIFOs introduced by the DBCF architecture,
a rejected LLR datum is pushed into a FIFO associated with the SISO
decoder that produced the rejected LLR datum.

\subsubsection{Further Analysis of DBCF Architecture}

\begin{figure}[t]
\begin{centering}
\includegraphics[width=0.7\columnwidth]{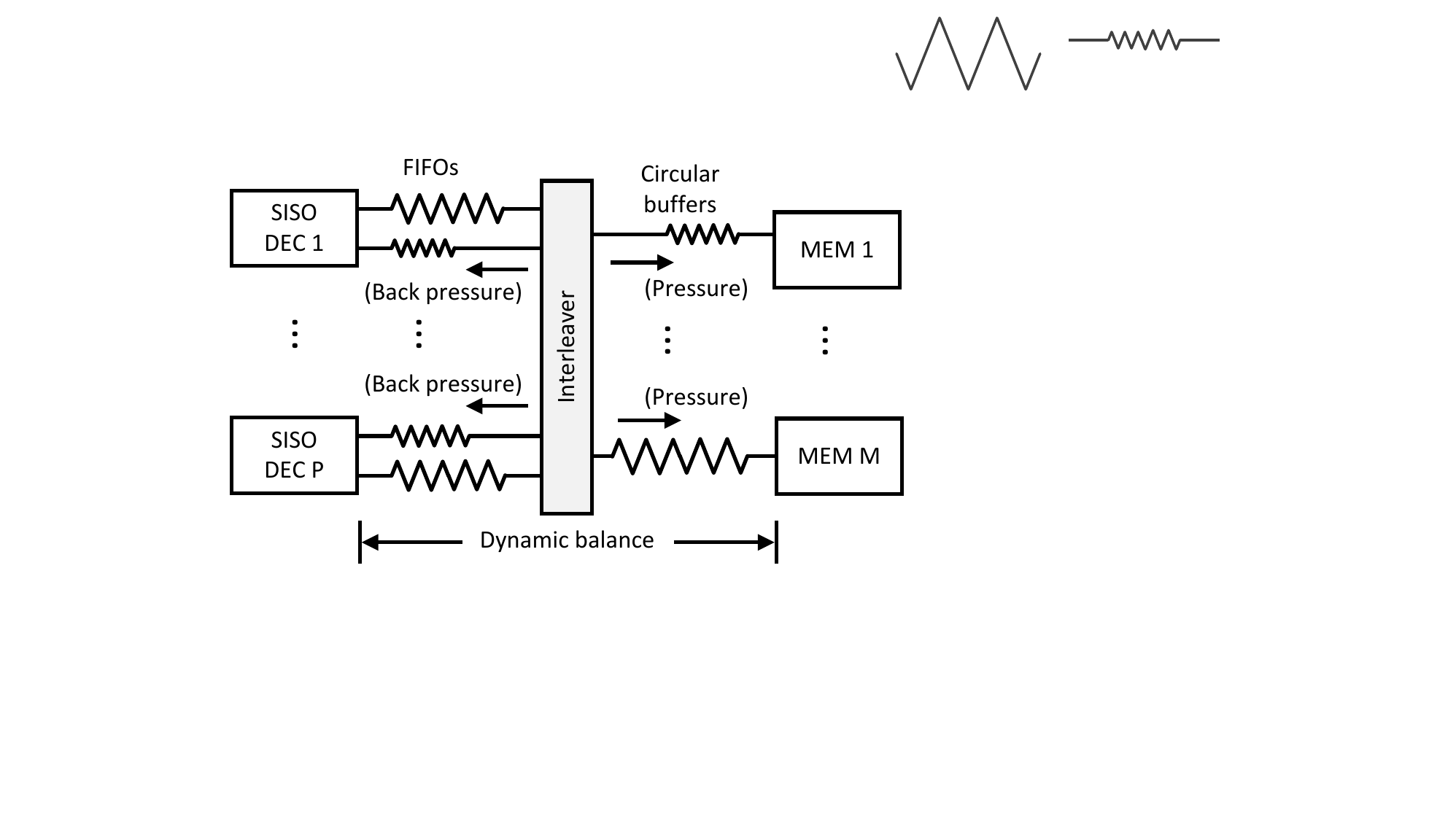}
\par\end{centering}

\vspace{-5pt}

\caption{Dynamic balance of data pressure on inter-connection network via ``spring-like''
structure of the DBCF architecture.\label{fig:spring_DBCF}}
\end{figure}

The DBCF architecture can significantly reduce the memory usage for
the data buffers. A model in Fig.~\ref{fig:spring_DBCF} can help
further understand the principle behind the DBCF architecture. The
two main components of the DBCF architecture, FIFOs and circular buffers,
work collaboratively like two sets of springs which distribute data
pressure evenly across both side of the inter-connection network.
Think about a possible case in which multiple SISO decoders send data
concurrently to memory-1 (MEM 1) in several consecutive clock cycles.
This burst behavior generates high data pressure on the circular buffer
associated with MEM 1. Therefore, if without FIFOs, a large circular
buffer is required to handle the occasional burst data accesses, which
leads to inefficient implementations. As a comparison, in DBCF architecture,
we use small circular buffers. Even when the circular buffer is fully
occupied by the bursty data, the incoming data are pushed into the
FIFOs, so that the circular buffers will not overflow and we do not
lose any data. That being said, the FIFOs and circular buffers together
maintain a dynamic balance of the data pressure on the interconnection
network. As a result, the size of the circular buffers can be significantly
reduced. Moreover, since the rejected data are stored in several FIFOs
associated with SISO decoders, with each FIFO accepting only one datum.
The data back pressure is gently distributed across FIFOs, so that
small FIFO size is sufficient.

In addition to the function of pressure balancing, the insertion of
FIFOs also decouples the SISO decoders from the interleaver. This
is important because of the following reasons: (1) with DBCF architecture,
SISO decoders can run at full speed without stalling, no matter how
severe the memory conflicts occur on the right hand side of interleaver.
This is a key to achieving high throughput; (2) any SISO decoder architecture
is supported to achieve high throughput. For instance, a Radix-4 XMAP
decoder generates 4 LLR data per clock cycle, among which some may
be successfully written into a memory module, while others may be
pushed back to FIFOs. But the Radix-4 XMAP decoder keeps producing
new outputs all the time. Without the help of the proposed double-buffer
architecture, a Radix-4 XMAP decoder has to stall even if only 1 LLR
value (out of 4) is rejected by a buffer router; (3) The decoupling
between SISO decoders and interleaver also reduces the complexity
of the SISO decoders since no feedback control from interleaver (or
memory modules) to the SISO decoders is needed.

\subsubsection{Determining Architectural Parameters}

\begin{table}[t]
\caption{Simulation results for parameter determination. \protect \\
HSPA+ interleaver, $K=5114$, $P_{LLR}=16$.\label{tab:Simulation-results-P16}}

\vspace{-5pt}

\footnotesize

\begin{centering}
\begin{tabular}{c|c|c|c|c|c|c|c}
\hline 
\multicolumn{5}{c|}{\textbf{Simulation parameters}} & \multicolumn{3}{c}{\textbf{Results}}\tabularnewline
\hline 
\multicolumn{1}{c|}{$P_{LLR}$} & \multicolumn{1}{c|}{$M$} & \multicolumn{1}{c|}{$S$} & $D_{FIFO}$ & $D_{buf}$ & \multicolumn{1}{c|}{$C_{0}$} & \multicolumn{1}{c|}{$C_{1}$} & \textbf{$\Delta C$}\tabularnewline
\hline 
\hline 
16 & 16 & 1 & 0 & 128 & 320 & 495 & 175\tabularnewline
\hline 
\hline 
16 & 16 & 3 & 4 & 12 & 320 & 332 & 12\tabularnewline
\hline 
16 & 32 & 3 & 3 & 4 & 320 & 323 & 3\tabularnewline
\hline 
\end{tabular}
\par\end{centering}

\normalsize
\end{table}

\begin{table}[t]
\caption{Simulation results for parameter determination. \protect \\
HSPA+ interleaver, $K=5114$, $P_{LLR}=32$.\label{tab:Simulation-results-P32}}

\vspace{-5pt}

\footnotesize

\begin{centering}
\begin{tabular}{c|c|c|c|c|c|c|c}
\hline 
\multicolumn{5}{c|}{\textbf{Simulation parameters}} & \multicolumn{3}{c}{\textbf{Results}}\tabularnewline
\hline 
\multicolumn{1}{c|}{$P_{LLR}$} & \multicolumn{1}{c|}{$M$} & \multicolumn{1}{c|}{$S$} & $D_{FIFO}$ & $D_{buf}$ & \multicolumn{1}{c|}{$C_{0}$} & \multicolumn{1}{c|}{$C_{1}$} & \textbf{$\Delta C$}\tabularnewline
\hline 
\hline 
32 & 32 & 1 & 0 & 128 & 160 & 268 & 108\tabularnewline
\hline 
\hline 
32 & 32 & 3 & 8 & 12 & 160 & 170 & 10\tabularnewline
\hline 
32 & 64 & 3 & 4 & 7 & 160 & 164 & 4\tabularnewline
\hline 
\end{tabular}
\par\end{centering}

\normalsize
\end{table}

\begin{figure*}[t]
\vspace{-10pt}
\begin{centering}
\includegraphics[width=0.7\textwidth]{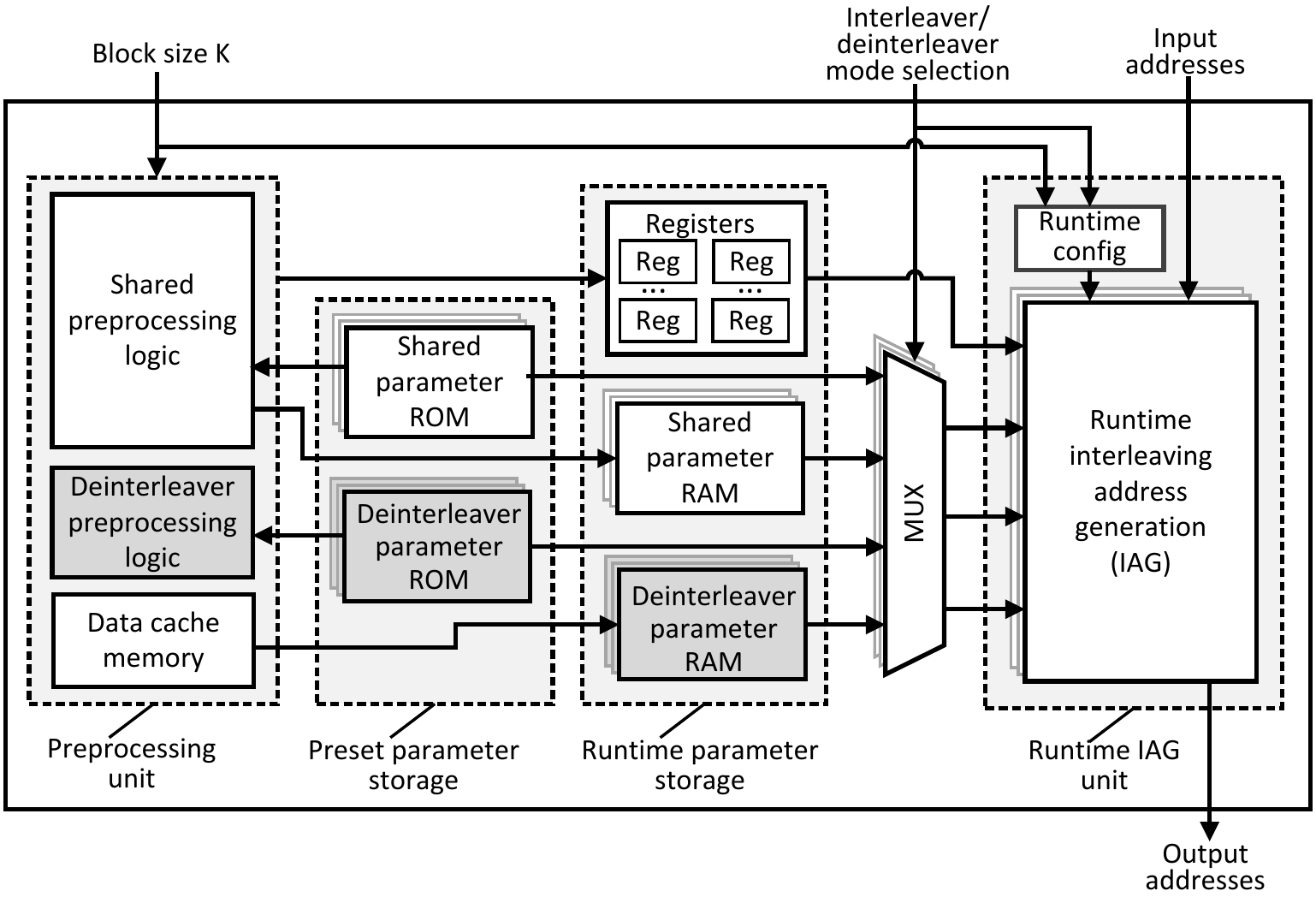}
\par\end{centering}

\vspace{-3pt}
\caption{Architecture of unified parallel interleaving address generation (IAG)
supporting both interleaving and deinterleaving. Note that we need
$P_{LLR}$ copies of each duplicated block in this figure. }

\label{Fig:arch}
\end{figure*}

To determine a parameter configuration (selection parameter $S$,
FIFO depth $D_{FIFO}$, circular buffer depth $D_{buf}$) for a parallel
turbo decoder with parallelism of $P_{LLR}$ and $M$ memory modules,
the cycle-accurate simulation is performed by adding the DBCF architecture
to the buffer system simulator. By changing different combinations
of ($S$, $D_{FIFO}$, $D_{buf}$), we can tune the design to meet
different performance/complexity goals in terms of memory usage, decoding
latency, and complexity of the interconnection network. In Table~\ref{tab:Simulation-results-P16}
and Table~\ref{tab:Simulation-results-P32}, parameter configurations
for $P_{LLR}=16$ and $P_{LLR}=32$ to achieve high throughput goal
as well as the corresponding simulation results are shown. Here, we
show results for the largest block size in HSPA+ standard ($K=5114$),
which typically causes more serious memory conflict problem and longer
latency. In these tables, $M$ denotes number of banks in the extrinsic
memories; $C_{0}$ denotes the number of clock cycles to decode a
codeword in an ideal case without any memory conflict; $C_{1}$ denotes
the actually clock cycles including the clock cycles used to solve
memory conflicts; $\Delta C$ denotes extra clock cycles actually
used compared to the ideal case ($\Delta C=C_{1}-C_{0}$). Although
all SISO decoders run at full speed without stalling in the proposed
architecture, the FIFOs and circular buffers still need a few clock
cycles to unload data, which causes extra clock cycles $\Delta C$.
A well-tuned parameter configuration can lead to small $\Delta C$.
Taking Table~\ref{tab:Simulation-results-P32} as an example, the
first row provides a baseline case in which only a large circular
buffer (without priority selection and FIFOs) is used to cache the
bursty data. In this baseline case, with buffer size set to 128 ($D_{buf}=128$),
108 extra clock cycles are needed to finish decoding. The second row
shows the usage of priority selection with $S=3$, $D_{FIFO}=8$ and
$D_{buf}=12$. In this case, 10 extra cycles are needed. The third
row shows the case in which sub-banking techniques (here, we set $M=2\times P_{LLR}$
for sub-banking) are used to further reduce $D_{FIFO}$ (reduced to
4) and $D_{buf}$ (reduced to 7), and only 4 extra clock cycles are
required to finish the decoding. Table~\ref{tab:Simulation-results-P16}
and Table~\ref{tab:Simulation-results-P32} have also demonstrated
the effectiveness and efficiency of the proposed DBCF architecture
in terms of the reduced buffer sizes and decreased extra clock cycles.

\section{Efficient Unified Interleaver/Deinterleaver Architecture\label{sec:Efficient-Unified-Interleaver/De}}

To improve the flexibility and configurability of a turbo decoder,
it is of great interest but also quite challenging to design and implement
an on-the-fly interleaving address generator (IAG) supporting both
interleaving and deinterleaving modes for multiple standards. In this
section, we propose a unified parallel IAG architecture to support
both interleaver and deinterleaver algorithms with low hardware overhead.

We notice that many interleaving laws in standards have similar computation
kernels for both interleaving and deinterleaving. For instance, a
pseudo-random column-row permutation algorithm is employed in 3GPP
UMTS/HSPA+ standards~\cite{3GPP_HSPA+}. The core operation of the
HSPA+ interleaver can be summarized as computing a column index function
$U_{i}(j)$ from the original column $j$ after intra-row permutation
$j\rightarrow U_{i}(j)$ for the $i$-th row, using the formula $U_{i}(j)=s((j\times r_{i})\mod(P-1))$.
Similarly, it has been proved that the core operation in the UMTS/HSPA+
deinterleaver is to compute an intra-row permutation $j'\rightarrow U{}_{i}^{-1}(j')$
using formula $U_{i}^{-1}(j')=(s^{-1}(j')\times m_{i})\mod(P-1)$~\cite{vosoughi:ASAP2013:UMTS_interleaver}.
The core computation units in the interleaver and deinterleaver can
be summarized as the same computation kernel $(a\times b)\mod c$.
Furthermore, most of the intermediate values precomputed by the pre-processing
units can be reused, except for only a few intermediate values used
exclusively by the deinterleaver mode. 

Similar observations can be made by investigating the LTE QPP (Quadratic
Permutation Polynomial) interleaving algorithm~\cite{3GPP_LTE}.
The QPP interleaver implements a permutation based on the quadratic
polynomial $f(x)=(f_{1}x+f_{2}x^{2})\mod K$. Efficient implementations
by recursively computing $f(x)$ have been proposed~\cite{sun2011efficient,Studer:2011:JSSC}.
Researchers have proved that the inverse permutation in the form of
quadratic polynomials exist for most of the $K$ (153 out of 188),
and for other $K$ inverse permutation in the forms of cubic (31/188)
or quartic (4/188) polynomials can be found~\cite{ryu:TIT2006:inverse_QPP,lahtonen:TIT2012:inverse_QPP}.
Similarly, the deinterleaving addresses can also be computed recursively
using efficient hardware. Therefore, the QPP interleaver and deinterleaver
can also share the kernel computation unit ($(a+bx)\mod c$).

Based on the above observations, we propose a unified interleaver/deinterleaver
architecture which shares a large portion of the hardware between
the interleaver and deinterleaver modes, leading to only a small overhead
compared to a single-mode interleaver. Fig.~\ref{Fig:arch} shows
the proposed unified parallel IAG architecture, which consists of
the following key blocks: (1) preset parameter storage, (2) preprocessing
unit, (3) runtime parameter storage, and (4) runtime IAG block. Here,
let us take the column-row random interleaving algorithm in the HSPA+
standard as an example to explain how these blocks work. The algorithm
details and notation definitions can be found from 3GPP UMTS/HSPA+
standards~\cite{3GPP_HSPA+} or references~\cite{Wang:2007:TCASII,vosoughi:ASAP2013:UMTS_interleaver}.
The preset parameter storage implemented using ROMs contains static
parameters such as inter-row permutation pattern $T_{i}$ ($128\times5bit$
tROM) and permutated sequence $r_{i}$ ($1440\times7bit$ rROM) used
by both modes, and the modular multiplicative inverse sequence $m_{i}$
($1200\times7bit$ mROM) used exclusively by the deinterleaver~\cite{vosoughi:ASAP2013:UMTS_interleaver}.
The preprocessing unit computes the parameters used in the runtime
IAG block. The preprocessing unit first generates several runtime
parameters based on the block size $K$, such as the number of rows/columns
($R$/$C$) of the permutation matrix, and then stores them in registers.
It also produces permutation patterns such as the base sequence for
intra-row permutations $s_{i}$ ($256\times8bit$ SRAM) and the corresponding
inverse base sequence used in the deinterleaver mode $s_{inv,i}$
($256\times8bit$ SRAM). Finally, the runtime IAG block, which consists
of computation and control logic, converts input addresses to the
interleaved/deinterleaved addresses. Based on the input block size
$K$ and interleaver/deinterleaver mode selection, the runtime IAG
can be configured to work in the corresponding mode. 

As is shown in Fig.~\ref{Fig:arch}, we need to duplicate some memory
and computation units for $P_{LLR}$ times to generate $P_{LLR}$
parallel addresses in one clock cycle. It is worth noting that we
only need one single copy of the preprocessing hardware. The balanced
scheduling scheme proposed in Section~\ref{sec:balanced-scheduling}
makes it possible that the parallel interleaver/deinterleaver can
function in a time-division multiplexing manner (different half iterations,
to be specific). Therefore, a significant amount of hardware resources
for two modes can be shared between two half iterations, which include
the preprocessing unit, kernel computation units inside the runtime
IAG, as well as ROMs and RAMs storing necessary parameter arrays.
The hardware sharing feature has been integrated in our turbo decoder
implementation presented in Section~\ref{sec:VLSI-Implementation-Results},
which can demonstrate the efficiency of the proposed unified IAG architecture.

\section{VLSI Implementation\label{sec:VLSI-Implementation-Results}}

\begin{figure*}[t]
\begin{centering}
\includegraphics[width=0.75\textwidth]{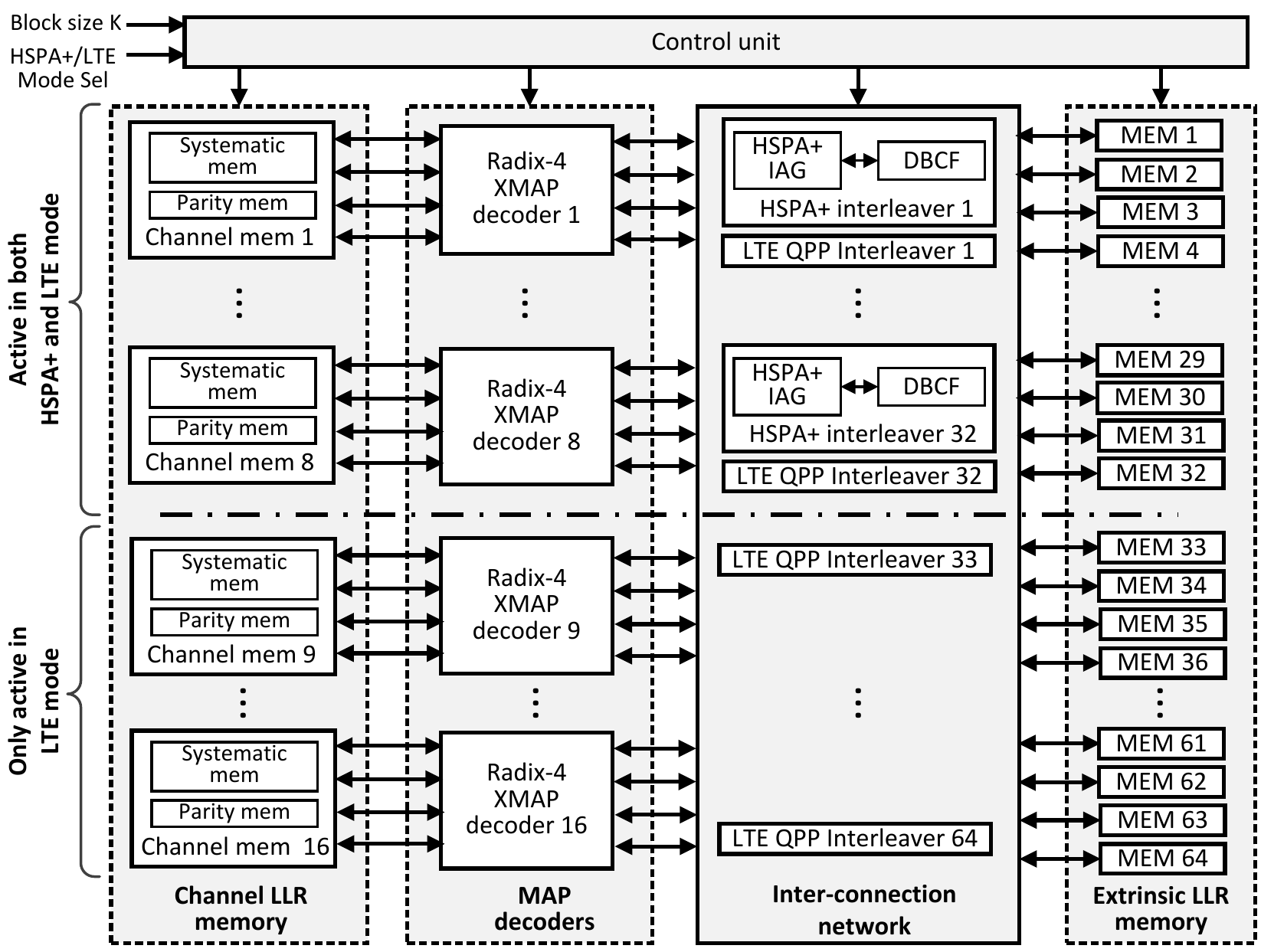}
\par\end{centering}

\caption{Overall architecture of the proposed multi-standard (HSPA+/LTE/LTE-Advanced)
turbo decoder.\label{fig:dual-mode-turbo}}
\end{figure*}

To show the effectiveness of the proposed architecture, we implement
a high throughput HSPA+/LTE multi-standard turbo decoder as is shown
in Fig.~\ref{fig:dual-mode-turbo}. The design goal is to achieve
more than 1\,Gbps throughput for LTE mode and 672\,Mbps throughput
for HSPA+ mode. 

\begin{figure}[t]
\begin{centering}
\includegraphics[width=1\columnwidth]{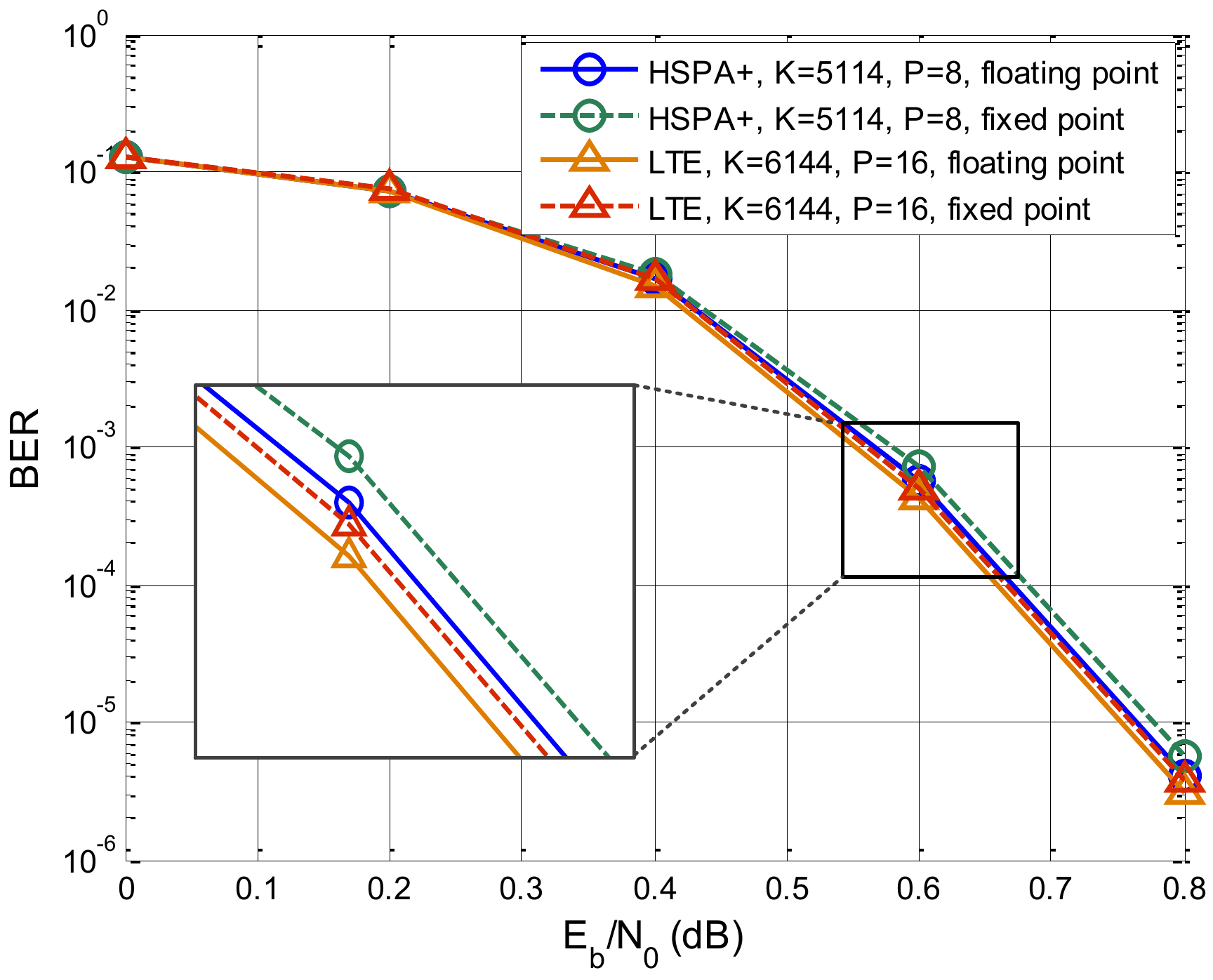}\vspace{-5pt}
\par\end{centering}

\caption{Bit error rate (BER) performance with fixed-point simulation for turbo
decoder (code rate 1/3) with the proposed parallel architecture, using
BPSK modulation over an additive white Gaussian noise (AWGN) channel.
The parallelism degrees for HSPA+ and LTE mode are 8 and 16, respectively. The number of
iterations is 5.5.
\label{fig:Fixed-point-simulation-results} }

\end{figure}

The Radix-4 XMAP decoding architecture described in Section~\ref{sub:Parallel-Turbo-Decoding}
is used to implement the SISO decoders. The max-log-MAP algorithm
is employed with a scaling factor 0.75 applied to improve the decoding
performance~\cite{vogt:EL2000:scaling_factor}. In the SISO decoder,
fixed-point representation is used for state metrics and LLR data
as follows: 5-bit channel LLR values, 6-bit extrinsic LLR values;
9-bit branch metrics ($\gamma$), and 10-bit state metrics ($\alpha$
and $\beta$). Fig.~\ref{fig:Fixed-point-simulation-results} shows
the fixed-point simulation results for both HSPA+ and LTE turbo decoder
using the proposed architecture and quantization. The simulation results
show that the fixed-point performance degradation is less than 0.1\,dB.

To achieve the throughput goal, the turbo decoder consists of $P=16$
Radix-4 XMAP decoders, which results in an effective parallelism of
64 ($P_{LLR}=P\times4=64$). Thus, the extrinsic LLR memory is partitioned
into 64 modules. Each extrinsic LLR memory module is implemented using
a $(K/P_{LLR}\times B_{ext})$ two-port SRAM ($B_{ext}$ is the word
length of the extrinsic LLR). Moreover, 16 channel LLR memory modules
are used to store the input channel LLR values. Each channel LLR memory
module is a $(3K/P\times2B_{ch})$ single-port SRAM ($B_{ch}$ is
the word length of the channel LLR). Every channel LLR memory module
is further partitioned into several banks to provide parallel accesses
to the systematic and parity check values. To support the high throughput
of LTE mode, all 16 SISO decoders work actively during decoding, with
the corresponding 64 extrinsic LLR memory modules, 16 channel LLR
memory modules, and 64 QPP interleavers. The HSPA+ interleaver and
DBCF buffer architecture will be bypassed under LTE mode. On the other
hand, when configured to work under the HSPA+ mode, the turbo decoder
uses 8 SISO decoders leading to an effective parallelism of $P_{LLR}=32.$
Accordingly, 32 extrinsic LLR memory modules, 8 channel LLR memory
modules, and 32 HSPA+ interleavers are employed. One DBCF buffer structure
per extrinsic LLR memory module is implemented to handle the memory
conflicts. The remaining 8 SISO decoders along with memory modules
are disabled using HSPA+ mode. 

The multi-standard HSPA+/LTE turbo decoder, which is described using
Verilog HDL, has been synthesized, placed and routed with the TSMC
45nm CMOS technology. In this section, we will describe the implementation
results in detail and comparisons with related work.

\subsection{Implementation Results for the Contention-free HSPA+ Interleaver\label{sub:hspa_intl_implementation_results}}

\begin{table}[t]
\scriptsize

\caption{Comparison with existing UMTS/HSPA+ interleavers. (Note: the chip
area includes IAG and interconnection-network, except for the first
three publications.)\label{tab:interleaver-comparison}}
\vspace{-5pt}

\begin{centering}
\begin{tabular}{c|c|c|c|c|c|c}
\hline 
\multirow{2}{*}{Work} & \multirow{2}{*}{$P_{LLR}$} & Flexible & \multicolumn{1}{c|}{Tech.} & Result & $Area$($A_{Norm}$$^{a}$) & $f_{clk}$\tabularnewline
 &  & $K$ & {[}nm{]} & type & {[}$mm^{2}${]} &  {[}MHz{]}\tabularnewline
\hline 
\hline 
\cite{Thul:2005:JVLSI} & 16 & N/A$^{b}$ & 180 & Syn. & 8.91 (0.254) & 166\tabularnewline
\hline 
\cite{Speziali:interleaver:DSD2004} & 32 & N/A$^{b}$ & 130 & Syn. & 2.784 (0.087) & 200\tabularnewline
\hline 
\cite{Neeb:noc:ISCAS:2005} & 16 & N/A$^{b}$ & 180 & Syn. & 1.2 (0.034) & 200\tabularnewline
\hline 
\cite{Ilnseher:2010:ICCS} & 16 & No & 40 & Syn. & 0.15 (0.086) & 350\tabularnewline
\hline 
\cite{Asghar_interleaver_JSPS_2012} & 2 & Yes & 65 & P\&R$^{c}$ & 0.014 (0.027) & 150\tabularnewline
\hline 
\cite{Wang:2007:TCASII} & 1 & Yes & 180 & Syn. & 0.24 (0.110) & 130\tabularnewline
\hline 
\cite{park:TVLSI2007:simd} & 1 & Yes & 250 & P\&R$^{c}$ & 2.69 (0.489) & N/A\tabularnewline
\hline 
\cite{Benkeser:2009:JSSC} & 1 & Yes & 130 & P\&R$^{c}$ & 0.4 (0.4) & 246\tabularnewline
\hline 
\multicolumn{1}{c|}{\cite{Borray:2009:ReConFig}} & \multicolumn{1}{c|}{1} & \multicolumn{1}{c|}{Yes} & FPGA  & Syn. & \multicolumn{1}{c|}{0.23$^{d}$ (0.229)} & \multicolumn{1}{c}{N/A}\tabularnewline
 &  &  & (90) &  &  & \tabularnewline
\hline 
\multicolumn{1}{c|}{\textbf{This}} & \multicolumn{1}{c|}{$32$} & \multicolumn{1}{c|}{Yes} & \multicolumn{1}{c|}{45} & P\&R$^{c}$ & 0.43 (0.121) & 600\tabularnewline
\textbf{work} &  &  &  &  &  & \tabularnewline
\hline 
\multicolumn{7}{l}{\emph{$^{a}$} Area is scaled to 130nm assuming $Area\thicksim1/s^{2}$
($s$ is feature size of }\tabularnewline
\multicolumn{7}{l}{technology nodes) and is normalized to equivalent area per LLR value.}\tabularnewline
\multicolumn{7}{l}{$^{b}$ These papers only present the interconnection-network to solve
memory}\tabularnewline
\multicolumn{7}{l}{conflicts. The IAG is not included/reported in the paper.}\tabularnewline
\multicolumn{7}{l}{$^{c}$ ``P\&R'' indicates that the reported results are after place
and route.}\tabularnewline
\multicolumn{7}{l}{$^{d}$ The chip area was not reported in the paper. The chip area
is estimated }\tabularnewline
\multicolumn{7}{l}{using the gate count reported in the paper.}\tabularnewline
\end{tabular}
\par\end{centering}

\normalsize
\end{table}

The HSPA+ interleaver is the most critical part in the proposed multi-standard
turbo decoder design. To achieve up to 672\,Mbps data rate under
HSPA+ mode, the turbo decoder employs an effective parallelism of
$P_{LLR}=32$, which can result in severe memory conflicts as indicated
by Fig.~\ref{fig:Memory-conflict-ratio}. Therefore, we first present
implementation results for the HSPA+ interleaver. 

The HSPA+ interleaver is implemented based on the balanced scheduling
scheme, the DBCF buffer architecture and the unified parallel interleaver/deinterleaver
architecture described in the previous sections. The block size $K$
can be configured at runtime. When a new block size $K$ is set, the
preprocessing unit re-calculates the runtime parameters. The interleaver
can work at both interleaver and deinterleaver modes based on a mode
selection signal.

As a reference design, we first implemented a single-function IAG
which only supports the interleaving algorithm. The chip area is 0.249\,$mm^{2}$
when synthesized with the TSMC 45nm CMOS technology and targeting
600\,MHz clock frequency. Then we implemented a dual-function IAG
supporting both the interleaving and deinterleaving algorithms of
the HSPA+ standard. The area for the unified IAG is 0.278\,$mm^{2}$.
Thus, compared to the single-function IAG, the chip area of the dual-function
IAG is only increased by 0.029\,$mm^{2}$, which indicates a 11.6\%
complexity overhead compared to a single-function IAG. This result
has demonstrated that the proposed unified parallel IAG architecture
enables great hardware sharing, leading to an efficient hardware implementation.
Inside the unified parallel IAG, the preprocessing logic utilizes
806\,$\mu m^{2}$ chip area, and the runtime computing logic utilizes
0.097\,$mm^{2}$ to generate 32 parallel interleaved (or deinterleaved)
addresses in one clock cycle. The unified HSPA+ IAG uses 65\,Kb memory
as runtime parameter storage to support parallelism degree of 32 and
different block sizes. 

\begin{table*}[t]
\caption{VLSI implementation comparison with existing high speed UMTS/HSPA+
and multi-mode turbo decoders.\label{tab:turbo-comparison}}

\footnotesize

\vspace{-8pt}

\begin{centering}
\begin{tabular}{l|c|c|c|c|c|c|c}
\hline 
Publication & \textbf{This work} & TVLSI'07\,\cite{park:TVLSI2007:simd} & TCAS-II'08\,\cite{martina:TCASII2008:UMTS-WiMax} & JSSC'09\,\cite{Benkeser:2009:JSSC} & JSPS'12\,\cite{Asghar_interleaver_JSPS_2012} & ICCS'10\,\cite{Ilnseher:2010:ICCS} & TVLSI'11\,\cite{lin:VLSI2011:multistandard}\tabularnewline
\hline 
\hline 
\multirow{2}{*}{Standard(s)} & HSPA+$^{a}$ & W-CDMA$^{a}$ & UMTS$^{a}$ & \multirow{2}{*}{HSDPA$^{a}$} & HSPA+$^{a}$/LTE & HSDPA$^{a}$ & HSDPA$^{a}$\tabularnewline
 & /LTE & /CDMA2000 & /WiMax &  & /WiMax/DVB & /LTE & /LTE/WiMax\tabularnewline
\hline 
\multirow{2}{*}{SISO decoder} & \multirow{2}{*}{Radix-4 XMAP} & \multirow{2}{*}{Radix-2 MAP} & \multirow{2}{*}{Radix-2 MAP} & \multirow{2}{*}{Radix-2 MAP} & \multirow{2}{*}{Radix-4 MAP} & Radix-2  & \multirow{2}{*}{Radix-4 MAP}\tabularnewline
 &  &  &  &  &  & monolithic MAP & \tabularnewline
\hline 
$\mathrm{N}_{SISO}$ / $\mathrm{N}_{LLR}$$^{b}$ & 16/64 & 1/1 & 4/4 & 1/1 & 4/8 & 1/16 & 5/10\tabularnewline
\hline 
\multirow{2}{*}{Block size $K$ {[}bits{]}} & \multirow{2}{*}{5114/6144} & \multirow{2}{*}{5114} & \multirow{2}{*}{5114/4800} & \multirow{2}{*}{5114} & 5114/6144 & \multirow{2}{*}{5114/6144} & 5114/\tabularnewline
 &  &  &  &  & 2400/12280 &  & 6144/4800\tabularnewline
\hline 
Technology & 45 nm & 180 nm & 130 nm & 130 nm & 65 nm & 40 nm & 130 nm\tabularnewline
\hline 
Supply Voltage $V_{DD}$ & 0.81 V & 1.8 V & N/A & 1.2 V & 1.1 V & 1.1 V & 1.2 V\tabularnewline
\hline 
Result type & P\&R & P\&R & Synthesis & P\&R & P\&R & Synthesis & P\&R\tabularnewline
\hline 
Clock frequency & 600 MHz & 111 MHz & 200 MHz & 246 MHz & 285 MHz & 350 MHz & 125 MHz\tabularnewline
\hline 
Core area & 2.43 $mm^{2}$ & 9.0 $mm^{2}$ & N/A & 1.2 $mm^{2}$ & 0.65 $mm^{2}$ & 1.46 $mm^{2}$ & 6.4 $mm^{2}$\tabularnewline
\hline 
Gate count & 1470 K & 34.4 K & 204 K & 44.1 K & N/A & N/A & N/A\tabularnewline
\hline 
Memory & 550 Kb  & 201 Kb & 148.6 Kb & 120 Kb & 214 Kb & N/A & 150 Kb\tabularnewline
\hline 
Max $\mathrm{N}_{iteration}$ & 5.5 & 6 & 8 & 5.5 & 6 & 6.5 & 5\tabularnewline
\hline 
\multirow{4}{*}{Throughput} & 826 Mbps  & \multicolumn{1}{c|}{} & 12 Mbps  & \multicolumn{1}{c|}{} & 49.5 Mbps & \multicolumn{1}{c|}{} & 23.9 Mbps \tabularnewline
 & (HSPA+) & 4.1 Mbps & (UMTS) & 20.2 Mbps & (HSPA+) & 300 Mbps & (HSDPA)\tabularnewline
\cline{2-2} \cline{4-4} \cline{6-6} \cline{8-8} 
 & 1.67 Gbps  & (W-CDMA) & 90.26 Mbps  & (HSDPA) & 173.3 Mbps & (HSDPA/LTE) & 105.6 Mbps \tabularnewline
 & (LTE) &  & (WiMax) &  & (LTE) &  & (LTE)\tabularnewline
\hline 
Power consumption & 870 mW & 292 mW & N/A & 61.5 mW & 570 mW & 452 mW & 374.75 mW\tabularnewline
\hline 
\hline 
Normalized area & \multicolumn{1}{c|}{5.62} & \multicolumn{1}{c|}{} & \multirow{4}{*}{N/A} & \multicolumn{1}{c|}{} & \multicolumn{1}{c|}{9.39} & \multicolumn{1}{c|}{} & \multicolumn{1}{c}{29.46}\tabularnewline
efficiency$^{cd}$  & (HSPA+) & 66.36 &  & 5.94 &  (HSPA+) & 12.22 & (HSDPA)\tabularnewline
\cline{2-2} \cline{6-6} \cline{8-8} 
{[}$\mathrm{mm^{2}}$/Mbps{]} & 2.77 & (W-CDMA) &  & (HSDPA) & 2.68 & (HSDPA/LTE) & 6.67\tabularnewline
\emph{(the lower, the better)} & (LTE) &  &  &  & (LTE) &  & (LTE)\tabularnewline
\hline 
Normalized energy & 0.55 & \multicolumn{1}{c|}{} & \multirow{4}{*}{N/A} & \multicolumn{1}{c|}{} & 3.84  & \multicolumn{1}{c|}{} & 3.14\tabularnewline
efficiency$^{c}$  & (HSPA+) & 8.57 &  & 0.55 & (HSPA+) & 0.75 &  (HSDPA)\tabularnewline
\cline{2-2} \cline{6-6} \cline{8-8} 
{[}nJ/bit/iter.{]} & 0.27  & (W-CDMA) &  & (HSDPA) & 1.10 & (HSDPA/LTE) & 0.71 \tabularnewline
\emph{(the lower, the better)} & (LTE) &  &  &  &  (LTE) &  & (LTE)\tabularnewline
\hline 
Architecture & 0.47 &  & \multirow{4}{*}{N/A} &  & 0.41 &  & 0.15\tabularnewline
efficiency$^{ce}$ & (HSPA+) & 0.05 &  & 0.38 & (HSPA+) & 0.42 &  (HSDPA)\tabularnewline
\cline{2-2} \cline{6-6} \cline{8-8} 
{[}bits/cycle/iter./$mm^{2}${]} & 0.96 & (W-CDMA) &  & (HSDPA) & 1.44 & (HSDPA/LTE) & 0.66\tabularnewline
\emph{(the higher, the better)} & (LTE) &  &  &  &  (LTE) &  & (LTE)\tabularnewline
\hline 
\multicolumn{8}{l}{$^{a}$ The W-CDMA, HSDPA and HSPA+ are terms referring to different
generations of technologies used in the 3GPP UMTS standard. }\tabularnewline
\multicolumn{8}{l}{\hspace{6pt}They all use the same turbo code structure and interleaving
algorithms.}\tabularnewline
\multicolumn{8}{l}{$^{b}$ $N_{LLR}$ denotes the number of parallel LLR values produced
per clock cycle, which represents the effective parallelism of a turbo
decoder. }\tabularnewline
\multicolumn{8}{l}{$^{c}$ Technology is scaled to 130nm CMOS assuming: $Area\thicksim1/s^{2}$,
and $t_{pd}\,(propagation\, delay)\thicksim1/s$. ($s$: feature size
of a technology node). }\tabularnewline
\multicolumn{8}{l}{\hspace{6pt}Throughput is linearly scaled to 5.5 iterations.}\tabularnewline
\multicolumn{8}{l}{$^{d}$ Normalized area efficiency = $Area_{Norm}/\ensuremath{Throughput_{Norm}}$~\cite{sun2011efficient,Studer:2011:JSSC,kienle:TCOM2011:turbo_decoder_complexity,lin:VLSI2011:multistandard}. }\tabularnewline
\multicolumn{8}{l}{$^{e}$ Architecture efficiency = $Throughput\times N_{iter}/Area_{Norm}/f_{clk}$~\cite{murugappa:DATE2013:turbo}.}\tabularnewline
\end{tabular}
\par\end{centering}

\normalsize
\end{table*}

Table~\ref{tab:interleaver-comparison} summarizes the key features
of the implemented contention-free HSPA+ interleaver, as well as the
comparison with related work. All the work shown in Table~\ref{tab:interleaver-comparison}
focus on the design of high performance and efficient interleavers
for the 3GPP UMTS/HSPA+ turbo decoder. Among them, references~\cite{Thul:2005:JVLSI,Speziali:interleaver:DSD2004,Neeb:noc:ISCAS:2005,Asghar_interleaver_JSPS_2012,Ilnseher:2010:ICCS}
concentrate on the interconnection-network solving the memory conflict
problem, so they did not report the area for the IAG modules. While
the others present efficient solutions for IAG modules for the UMTS/HSPA+.
To make a fair comparison, we show the normalized chip area for each
implementation; furthermore, for those supporting parallel interleaving
address generation, the chip area is further normalized by the supported
parallelism to determine an equivalent chip area for each generated
interleaving address. The results in \cite{Wang:2007:TCASII,park:TVLSI2007:simd,Borray:2009:ReConFig,Benkeser:2009:JSSC}
support on-the-fly interleaving address generation for HSPA+, but
they do not support parallel decoding due to the lack of a memory
conflict solver. The interleavers in \cite{Thul:2005:JVLSI,Ilnseher:2010:ICCS}
are not reconfigurable because they support only one block size. Only
\cite{Ilnseher:2010:ICCS,Borray:2009:ReConFig} and our work support
both the interleaver and deinterleaver modes. The architecture in
\cite{Asghar_interleaver_JSPS_2012} can support a parallel turbo
decoder with different block sizes, but the architecture is limited
to turbo decoders with low parallelism. Our proposed architecture
shows the best tradeoff among flexibility, configurability, and hardware
complexity.

\subsection{Implementation Results of Multi-Standard Turbo Decoder}

\begin{figure}[t]
\begin{centering}
\includegraphics[width=0.9\columnwidth]{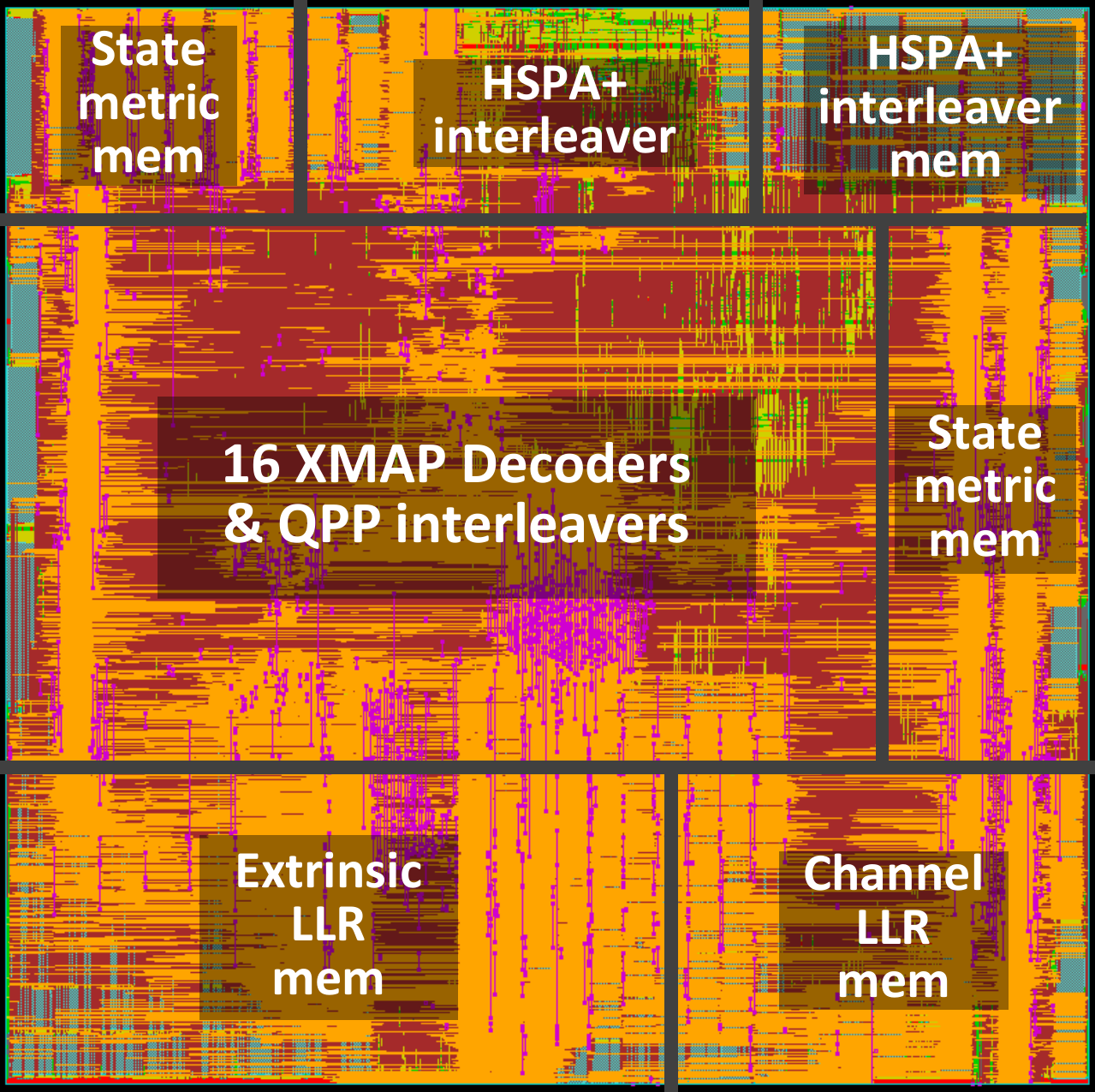}
\par\end{centering}

\caption{VLSI chip layout for an HSPA+/LTE multi-standard turbo decoder with
the architecture shown in Fig.~\ref{fig:dual-mode-turbo}. The TSMC
45nm standard cell library is used. This turbo decoder contains 16
Radix-4 XMAP decoders, 64 LTE QPP interleavers, and 32 HSPA+ interleavers.\label{fig:layout_view}}
\end{figure}

The key characteristics of the implemented multi-standard turbo decoder
are summarized in Table~\ref{tab:turbo-comparison}. The design is
synthesized using the TSMC 45nm standard cell library. The chip core
area of the implemented turbo decoder is 2.43\,$mm^{2}$. A top-level
chip layout view after place and route is shown in Fig.~\ref{fig:layout_view}.
The maximum number of iterations of this turbo decoder is set to 5.5.
The block size $K$ can be configured at runtime ($K=40\sim5114$
for HSPA+; $K=40\sim6144$ for LTE). When clocked at 600\,MHz, a
maximum throughput of 1.67\,Gbps can be achieved under the LTE mode
with all 16 Radix-4 XMAP decoders running. While in the HSPA+ mode,
a maximum throughput of 826\,Mbps can be achieved with 8 Radix-4
XMAP decoders active. Compared to a single-mode LTE turbo decoder,
the area overhead of the HSPA+/LTE multi-standard turbo decoder is
approximately 17\% in our implementation ($0.426mm^{2}/2.43mm^{2}$). 

To the best of our knowledge, our multi-standard turbo decoder implementation
is the first to support the peak data rate of 336\,Mbps in Release
11 of the 3GPP UMTS/HSPA+ standard~\cite{3GPP_HSPA+}. Moreover,
our implementation also exceeds the 672\,Mbps peak data rate requirement
of the potential future extension of the HSPA+ standard and the 1\,Gbps
peak data rate requirement for the LTE-Advanced standard~\cite{3GPP_LTE}. 

A comparison with related work can be found in Table~\ref{tab:turbo-comparison}.
These related publications are either high speed UMTS/HSPA+ turbo
decoder designs or multi-standard turbo decoders supporting the UMTS/HSPA+
standard. Implementation details such as clock frequency, chip core
area, throughput, and power consumption are included. We adopt three
normalized metrics to evaluate implementation efficiency: area efficiency,
energy efficiency and architecture efficiency~\cite{Studer:2011:JSSC,sun2011efficient,kienle:TCOM2011:turbo_decoder_complexity,lin:VLSI2011:multistandard,murugappa:DATE2013:turbo}.
The normalization is done by scaling the reported throughput and area
numbers to 130nm CMOS technology. From Table~\ref{tab:turbo-comparison},
we can see that our implementation shows the best normalized energy
efficiency among the publications under both HSPA+ and LTE mode. In
HSPA+ mode, our implementation shows the best area (or architecture)
efficiency. While for LTE mode, our implementation has better normalized
area (or architecture) efficiency than most of papers listed in Table~\ref{tab:turbo-comparison}.
Note that only \cite{Asghar_interleaver_JSPS_2012} shows better area
(or architecture) efficiency. However, it has lower energy efficiency
than the proposed implementation. Moreover, the parallel interleaver
architecture proposed by \cite{Asghar_interleaver_JSPS_2012} only
supports low parallelism for the HSPA+ mode, which limits its achievable
throughput under HSPA+ mode.

\section{Conclusion\label{sec:Conclusion}}

In this paper, we propose a VLSI architecture for highly parallel
turbo decoding aiming at multi-standard 3G/4G wireless communication
systems. We first utilize the balanced scheduling scheme to avoid
memory reading conflicts. Then, based on the statistical property
of memory conflicts, we propose a double-buffer contention-free (DBCF)
buffer architecture to eliminate memory writing conflicts. Furthermore,
we propose an efficient unified parallel interleaver architecture
supporting both interleaving and deinterleaving modes. It is worth
mentioning that the proposed parallel interleaver architecture is
independent from the SISO decoder architectures, the interleaver algorithms,
and parallelism degrees, thus, providing good flexibility and scalability.
To demonstrate the effectiveness of our design approaches, we synthesize,
place and route an ASIC design for the HSPA+/LTE/LTE-Advanced multi-standard
turbo decoder using the proposed VLSI architecture with a 45nm CMOS
technology. The chip core area is a 2.43\,$mm^{2}$. When clocked
at 600\,MHz, this turbo decoder can achieve a maximum throughput
of 826~Mbps in HSPA+ mode, and a peak throughput of 1.67\,Gbps in
LTE/LTE-Advanced mode. The implementation results show that as an
enabler, the proposed parallel interleaver architecture is able to
lead to highly parallel but also hardware-efficient turbo decoding
implementations.

\section*{}

\footnotesize

% Generated by IEEEtran.bst, version: 1.13 (2008/09/30)

\end{document}